\documentclass[conference]{IEEEtran}
\usepackage[T1]{fontenc}
\usepackage{booktabs}
\usepackage{graphicx}
\usepackage{siunitx}
\usepackage{tikz}
\usepackage{pgfplots}
\pgfplotsset{compat=1.17}
\usepgfplotslibrary{groupplots}
\usepackage[hidelinks]{hyperref}
\usepackage{amsmath}
\usepackage{array}
\newcolumntype{P}[1]{>{\raggedright\arraybackslash}p{#1}}
\usepackage{url}
\makeatletter\def\input@path{{figures/}}\makeatother
\usepackage{hades-tikz}

\newcommand{\hades}{\textsc{HADES}}
\newcommand{\isa}[1]{\texttt{\detokenize{#1}}}
\title{HADES: A Programmable Processing-Near-Memory Engine for Binary Similarity Search}
\author{\IEEEauthorblockN{Pranav M.}
\IEEEauthorblockA{C.H.I.P.S. Lab, PES University\\
pranav.m1205@gmail.com}}

\begin{document}
\maketitle

\begin{quote}\small\noindent\textbf{\textit{Abstract:}}
Moving operands to a processor can cost more than the arithmetic performed on them, and for small binary similarity and search workloads data movement can dominate. We present \hades{}, a programmable processing-near-memory engine built around one shared $32\times32$-bit 2R1W register-file macro that holds both data and program. Twelve 16-bit opcodes, three working registers and a 32-slot program express ten search workloads in 8 to 19 slots, and \isa{SCANP} and \isa{SCANSEL} shorten dual-row and selected scans. A post-layout $2\times2$ Tiny Tapeout implementation closes timing at $40$\,MHz with $26{,}968$\,\textmu m$^2$ of standard cells beside a $15{,}744$\,\textmu m$^2$ macro. All \hades{} results come from RTL or gate-level simulation of this implementation, and \hades{} energy is estimated. Against ESP8266 and ESP32 microcontrollers running plain C, and with an ideal host, \hades{} answers a resident query $10\times$ to $32\times$ and $2.2\times$ to $11\times$ faster, respectively. The microcontroller per-query times include record serialisation and a running correctness hash, so these factors are upper bounds; the compute-only time of the same engine gives lower bounds of $22.5\times$ and $7.1\times$ for nearest-neighbour and equality on the ESP8266 and $8.9\times$ and $2.8\times$ on the ESP32. End to end, \hades{} has the lower total at the evaluated batch sizes of $B=16$ and $B=256$ against both boards, and loses every single-query ($B=1$) run to the ESP32 and eight of the twenty to the ESP8266. An Arduino Nano runs a restricted subset. Kernel power stays within about $2\%$ of a spin loop on all three boards. These results hold for plain-C baselines under the evaluated conditions. At this scale the control and instruction storage fit around a shared memory macro, while the byte-wide interface and the loading path remain significant costs.
\end{quote}

\section{Introduction}
\label{sec:intro}

In most computing systems, moving the operands to the processor dominates the cost of an operation on large data. Data movement between memory and compute, especially off-chip to on-chip, is much more expensive in latency, energy and bandwidth than computation, and memory bandwidth and energy scale poorly as applications become more data-intensive~\cite{mutlu2020primer}. DAMOV characterises where such bottlenecks arise in real applications~\cite{oliveira2021damov}.

Processing-near-memory (PNM) places logic close to the memory, so that a reduction over resident data runs where the data lives and only a small result crosses the memory interface~\cite{mutlu2020primer}. A compute-near-memory unit of this kind leaves the memory cell arrays unmodified~\cite{medina2025bank}. Published PNM and processing-in-memory designs include commercial DRAM-resident processors~\cite{gomezluna2021upmem}, accelerators for graph processing~\cite{ahn2015tesseract}, programmable near-memory units for embedded systems~\cite{caon2024nmc,petrolo2025arcane} and bank-level compute for DRAM~\cite{medina2025sidedram}. Most of these designs target matrix, neural-network or graph kernels.

Binary similarity and search over short bit-vector signatures sits at the opposite end of this design space. The datapath is a masked XOR followed by a population count, and the output is a small record: a distance and an index, a match bitmap, a count. A scan must read every resident row, yet its answer is a few bytes. The compute is nearly free relative to the memory it reads, so the workload has the shape a near-memory engine exploits: the whole dataset is read and only a short record is returned. In the plain-C baselines we measure on three microcontrollers, the compiler implements popcount as a library call.

If the datapath is this small, the cost that remains to be justified is the cost of making the engine \emph{programmable}: instruction storage, sequencing and working registers. The workload class contains many kernels (nearest-neighbour, range, exact match, population analytics, filter-then-refine) that share one datapath and differ in control flow, so a fixed-function unit would cover only one of them. We state the question as follows.

\begin{quote}
\emph{What programmable substrate (ISA width, instruction depth, control overhead) is sufficient to express useful binary similarity and search workloads against locally resident memory, and what physical area and execution costs result?}
\end{quote}

We answer with \hades{} (Hamming-Aware Decode, Execute, and Search), a small, stored-program PNM engine. It uses an unmodified third-party register-file macro through the macro's ordinary read and write ports. Around that macro it adds a 16-bit instruction set of twelve opcodes, three working registers, a small set of result registers, and a 32-slot program held in the same macro as the data. \hades{} performs processing \emph{near} memory: the memory cells are neither modified nor used to compute. Besides the basic scan instruction, the ISA includes two scan variants that target the two structural limits of a one-row-per-cycle scan. \isa{SCANP} uses the macro's otherwise idle second read port to scan two rows per cycle for the operations that need no popcount. \isa{SCANSEL} restricts a scan to the rows selected by a previous predicate scan, so a filter-then-refine query stays inside the engine. The design has been hardened through the Tiny Tapeout~\cite{tinytapeout} physical-design flow on a $2\times2$ tile.

The evaluation compares \hades{} with plain C on three microcontrollers at three measurement boundaries and shows where \hades{} wins and where it does not. The evidence is post-layout and gate-level for \hades{}, measured for the microcontroller baselines, and estimated for \hades{} energy. The compact programmable engine expresses all ten workloads, and it is faster than both ESP boards once the dataset and program are resident, although the baseline per-query times include record hashing that the \hades{} boundary does not (Section~\ref{sec:br}). Loading the dataset and program through a byte-wide interface decides the end-to-end outcome.

The contributions are the following.
\begin{itemize}
\item An architecture and instruction set for a small programmable near-memory engine for binary similarity search, built around a single shared register-file macro that holds both data rows and program (Section~\ref{sec:arch}).
\item Two small architectural mechanisms, \isa{SCANP} and \isa{SCANSEL}, that exploit the macro's second read port and an in-memory selection bitmap, and a cycle model for every instruction (Section~\ref{sec:arch}).
\item A set of ten workload definitions with canonical result records and program templates, and a measure of how many of the 32 program slots each template consumes (Section~\ref{sec:workloads}).
\item A hardened post-layout implementation on a $2\times2$ tile, and an evaluation that separates compute-only, resident-query and end-to-end measurement boundaries and compares against ESP8266 and ESP32 microcontrollers, and against an Arduino Nano on a restricted subset, under byte-exact result checking (Sections~\ref{sec:impl} and~\ref{sec:eval}).
\end{itemize}
Each result is labelled by how it was obtained (RTL or gate-level simulation, measurement on a commodity platform, model, or estimate), and numbers taken at different boundaries are never combined.

Section~\ref{sec:related} reviews related work, and Section~\ref{sec:motivation} motivates PNM for binary similarity search. Section~\ref{sec:arch} describes the architecture, and Section~\ref{sec:workloads} the workloads it expresses. Section~\ref{sec:impl} covers the implementation and Section~\ref{sec:eval} the evaluation. Discussion, limitations and conclusions follow.

\section{Related Work}
\label{sec:related}

\emph{PIM and PNM characterisation.} A broad primer reviews processing-in-memory and processing-near-memory~\cite{mutlu2020primer}, and DAMOV provides a methodology and benchmark suite for locating data-movement bottlenecks in applications~\cite{oliveira2021damov}. Both stress matching an architecture to a workload's bottleneck, which is the approach \hades{} takes for one narrow workload class. \hades{} does not target the application scale they consider.

\emph{Programmable PIM systems.} The UPMEM system attaches a small programmable processor to each DRAM bank. Benchmarking identifies streaming, low-synchronisation, simple-arithmetic workloads as its best fit~\cite{gomezluna2021upmem}. Tesseract is an in-memory accelerator for parallel graph processing, and PIM-enabled instructions provide locality-aware offload of individual instructions to memory~\cite{ahn2015tesseract,ahn2015pei}. These systems establish that programmable logic next to memory pays off for streaming work. Binary similarity search shares the streaming, simple-arithmetic profile. The gap \hades{} addresses is how much control and storage a single register-file-scale engine needs.

\emph{Near-memory computing for embedded and edge systems.} Two programmable near-memory computing architectures on SRAM for microcontroller-class nodes trade area, performance and flexibility against one another. One is controlled directly by the host, and the other adds a small embedded RISC-V controller for autonomous execution~\cite{caon2024nmc}. ARCANE extends a RISC-V cache so that custom host instructions dispatch vector operations to near-memory units~\cite{petrolo2025arcane}. These designs target dense linear-algebra and neural-network kernels with vector datapaths. \hades{} resembles them in using conventional memory macros and in asking how much control is justified. It differs in scale, in workload, and in being a complete program-driven engine the size of a single register file.

\emph{Bank-level compute and design-space exploration.} SideDRAM integrates SoftSIMD datapaths and a distributed control plane near DRAM banks~\cite{medina2025sidedram}. Exploration frameworks exist for near-bank and near-memory designs~\cite{medina2025bank,kodra2026nmc}. One reports that the best ratio between instruction and data capacity varies between roughly $2\times$ and $4\times$ across the DRAM-bank processing units and benchmarks it studies~\cite{medina2025bank}. We cite this as evidence that instruction-storage capacity is a sizing variable in near-memory units. \hades{} is a much smaller design point. Its data and program capacities are fixed by the macro and by the row-index and program-counter widths, not derived from a sizing study of this kind (Section~\ref{sec:arch}).

\section{Motivation: Processing-Near-Memory}
\label{sec:motivation}

\subsection{Bandwidth gap and data reduction at the host interface}

\textbf{Key insight: a scan reads everything and returns almost nothing.} The argument for PNM is structural. A scan over $N$ resident rows must read every row, but its answer is usually far smaller than the data it summarises. When the compute unit sits behind the same interface as the host, every read crosses that interface. When it sits next to the memory, only the answer crosses. This is the principle behind the move from processor-centric to data-centric designs~\cite{mutlu2020primer,oliveira2021damov}. Whether an operation should run near memory also depends on data locality, because data already held close to the host is cheaper to reuse there~\cite{ahn2015pei}.

The register-file macro reads up to $32$ bits per read port per cycle, and a dual-row scan uses both read ports. The chip's interface carries at most one input byte and one output byte per cycle, because the Tiny Tapeout platform provides eight dedicated input and eight dedicated output pins~\cite{tinytapeout}. A $16$-row dataset occupies $64$ bytes. A nearest-neighbour scan summarises it in a two-byte record, an equality scan in three, and a threshold scan in six (Section~\ref{sec:workloads}). As a hypothetical comparison, a host that read the same $64$ bytes through an eight-bit-wide path would need at least $64$ transfer cycles, whereas \hades{} returns a record of two to six bytes. This arithmetic uses widths and record sizes only, and no \hades{} transfer was measured.

The ratios bound the possible saving, because \hades{} has no instruction that streams data rows out and its dataset must itself be loaded through a byte-wide path. The engine also pays costs that a host-side scan avoids: the query must be delivered, the program and dataset must be loaded, and the specified clock is $40$\,MHz. Section~\ref{sec:eval} tests whether the structural reduction survives these costs by separating compute-only, resident-query and end-to-end boundaries.

\subsection{Why binary similarity search}

Binary similarity and search is a good match for a small programmable engine for three reasons.

\textbf{A small, uniform datapath.} With $D_i$ denoting the $i$th stored row, $Q$ a query and $M$ a mask, the basic quantity is the masked Hamming distance $d_i = \operatorname{pop}((D_i \oplus Q)\wedge M)$. The other operations of interest (equality, thresholded range search, masked population counts and bitwise reductions) are minor variations on it. There is no multiplication and no wide accumulation, so the datapath is small and the engine's size is set largely by what surrounds it.

\textbf{Small, structured results.} The results are a best index and distance, a bitmap of matching rows with a count, or a reduced word. A scan therefore ends in a short record of a few output bytes, which gives data reduction at the interface its meaning.

\textbf{Many kernels on one datapath.} Nearest-neighbour, range and exact-match queries, feature selection by mask, population analytics over the stored rows, and a filter-then-refine combination share the same datapath yet differ in control flow and in the result registers they use. That variety is what makes a \emph{programmable} substrate a meaningful question rather than a fixed-function unit.

The workloads in Section~\ref{sec:workloads} are defined by their semantics, independent of any particular application, so that every platform in the evaluation executes the same function. The scale is deliberately modest. A row is $32$ bits and the engine holds $16$ rows, so the design does not aim at vector-database scale, and the host handles datasets larger than $16$ rows by tiling (Section~\ref{sec:workloads}). At this scale every resource is countable, which keeps the substrate question tractable.

\section{HADES Architecture}
\label{sec:arch}

\subsection{Overview and memory organisation}

\begin{figure}[t]
\centering
\begin{tikzpicture}[hades, x=1cm, y=1cm,
    core/.append style={font=\scriptsize}, reg/.append style={font=\scriptsize}, io/.append style={font=\scriptsize},
    mem/.append style={font=\scriptsize}]
  \node[io, minimum width=1.5cm, minimum height=0.5cm] (hld) at (0.76,6.05) {Loader\\[-2pt]{\tiny LD\_STROBE}};
  \node[io, minimum width=1.5cm, minimum height=0.5cm] (hex) at (0.76,5.45) {EXECUTE};
  \node[io, minimum width=1.5cm, minimum height=0.5cm] (hst) at (0.76,4.30) {READY\\[-2pt]{\tiny BUSY, HALT}};
  \node[io, minimum width=1.5cm, minimum height=0.5cm] (hrv) at (0.76,2.90) {RECV\\[-2pt]{\tiny ui, STROBE}};
  \node[io, minimum width=1.5cm, minimum height=0.5cm] (hsd) at (0.76,0.30) {SEND\\[-2pt]{\tiny uo[7:0]}};

  \node[core, minimum width=2.2cm, minimum height=0.7cm] (ldr) at (3.23,5.75) {Loader / EXECUTE\\[-2pt]{\tiny load, restart}};

  \node[core, minimum width=2.2cm, minimum height=0.95cm] (ctl) at (3.23,4.30) {Control\\[-2pt]{\tiny FSM, PC, sequencer}\\[-2pt]{\tiny\itshape SCANSEL: MATCH rows}};
  \node[reg,  minimum width=2.2cm, minimum height=0.7cm] (reg) at (3.23,2.90) {Operand regs\\[-2pt]{\tiny Q, A, MASK}};
  \node[core, minimum width=2.2cm, minimum height=0.95cm] (cmp) at (3.23,1.65) {Compute\\[-2pt]{\tiny XOR, AND, popcount}\\[-2pt]{\tiny\itshape SCANP: 2 rows / cycle}};
  \node[reg,  minimum width=2.2cm, minimum height=0.7cm] (res) at (3.23,0.30) {Result regs\\[-2pt]{\tiny best/next, MATCH, count}};
  \node[mux] (arb) at (5.22,2.50) {};

  \begin{scope}[on background layer]
    \node[group, fit=(reg)(cmp)(res), inner sep=3pt] (dp) {};
    \coordinate (cpad) at ([yshift=-0.22cm]dp.south);      
    \node[group, fit=(ctl)(dp)(arb)(cpad), inner sep=4pt] (core) {};
  \end{scope}
  \node[grouplabel, anchor=west] at ([xshift=4pt]core.south west) {HADES core};
  \node[grouplabel, anchor=east] at ([xshift=-4pt]dp.south east) {datapath};

  \node[mem, minimum width=2.0cm, minimum height=5.65cm, anchor=south west] (rf) at (6.0,-0.10) {};
  \node[anchor=north, font=\scriptsize\bfseries, align=center] at ([yshift=-2pt]rf.north) {rf\_top\\[-2pt]{\tiny\mdseries 2R1W, 32 $\times$ 32 b}};
  \node[reg, minimum width=1.75cm, minimum height=1.8cm] (rfp) at (7.0,3.85) {};
  \node[font=\tiny, align=center, anchor=north] at ([yshift=-2pt]rfp.north) {\textbf{Program}\\[-1pt]words 16--31\\[-1pt]16 $\times$ 32 b};
  \node[draw, line width=0.4pt, minimum width=0.78cm, minimum height=0.36cm, font=\tiny, inner sep=1pt, fill=white] at (6.61,3.50) {$2k{+}1$};
  \node[draw, line width=0.4pt, minimum width=0.78cm, minimum height=0.36cm, font=\tiny, inner sep=1pt, fill=white] at (7.39,3.50) {$2k$};
  \node[font=\tiny, anchor=north] at (7.0,3.30) {2 slots / word};
  \node[reg, minimum width=1.75cm, minimum height=1.4cm, font=\tiny, align=center] (rfd) at (7.0,1.20) {\textbf{Data rows}\\[-1pt]words 0--15\\[-1pt]16 $\times$ 32 b};

  \begin{scope}[on background layer]
    \node[group, fit=(ldr)(core)(rf), inner sep=4pt] (wrap) {};
  \end{scope}
  \node[grouplabel, anchor=east] at ([xshift=-4pt]wrap.north east) {tt\_um\_hades (wrapper)};

  \draw[data] (hld.east) -- (hld.east -| ldr.west);
  \draw[ctl]  (hex.east) -- (hex.east -| ldr.west);
  \draw[data] (hrv.east) -- (reg.west);
  \draw[data] (res.west) -- (hsd.east);
  \draw[ctl]  (ctl.west) -- (hst.east);
  \draw[ctl]  (ldr.south) -- (ctl.north);
  \node[font=\tiny, inner sep=1pt, anchor=west] at ([xshift=0.06cm]$(ldr.south)!0.5!(ctl.north)$) {exec};

  \draw[ctl] ([xshift=-0.6cm]ctl.south) -- ([xshift=-0.6cm]dp.north);
  \draw[ctl] ([xshift=0.6cm]dp.north) -- ([xshift=0.6cm]ctl.south);
  \node[font=\tiny, inner sep=1pt, fill=white, fill opacity=1, text opacity=1] at ([xshift=-0.6cm]$(ctl.south)!0.5!(dp.north)$) {scan op};
  \node[font=\tiny, inner sep=1pt, fill=white, fill opacity=1, text opacity=1] at ([xshift=0.6cm]$(ctl.south)!0.5!(dp.north)$) {MATCH};
  \draw[data] (reg.south) -- (cmp.north);
  \draw[data] (cmp.south) -- (res.north);

  \draw[bus, <->] (ctl.east) -- (ctl.east -| rf.west) node[wirelabel, pos=0.78, font=\tiny] {A};
  \draw[bus, <-]  (cmp.east) -- (cmp.east -| rf.west) node[wirelabel, pos=0.80, font=\tiny] {B};

  \coordinate (arbIn1) at ([yshift=0.20cm]arb.west);     
  \coordinate (arbIn2) at ([yshift=-0.20cm]arb.west);    
  \draw[data] (ldr.east) -- ([xshift=0.30cm]ldr.east) -- ([xshift=0.30cm]ldr.east |- arbIn1) -- (arbIn1);
  \draw[data] (reg.east) -- ([xshift=0.20cm]reg.east) -- ([xshift=0.20cm]reg.east |- arbIn2) -- (arbIn2);
  \draw[bus]  (arb.east) -- (arb.east -| rf.west) node[wirelabel, pos=0.5, font=\tiny] {W};
  \node[wirelabel, font=\tiny, align=center] at ([xshift=0.30cm,yshift=-0.55cm]ldr.east) {init,\\[-2pt]program};
  \node[font=\tiny, inner sep=1pt, fill=white, fill opacity=1, text opacity=1] at ($([xshift=0.20cm]reg.east)!0.5!([xshift=0.20cm]reg.east |- arbIn2)$) {STA};

  \draw[data] (0.6,-1.0)  -- ++(0.45,0) node[right, font=\tiny] {data};
  \draw[ctl]  (2.0,-1.0)  -- ++(0.45,0) node[right, font=\tiny] {control / status};
  \draw[bus]  (4.6,-1.0)  -- ++(0.45,0) node[right, font=\tiny] {32-bit RF port};
\end{tikzpicture}
\caption{High-level \hades{} architecture. The wrapper (outer dashed box) contains the host-facing loader and \isa{EXECUTE} detector; the core contains control, operand registers, compute, result registers and a write arbiter. The core reaches memory only through the three ports of \texttt{rf\_top}: read ports~A and~B and write port~W.}
\label{fig:arch-top}
\end{figure}
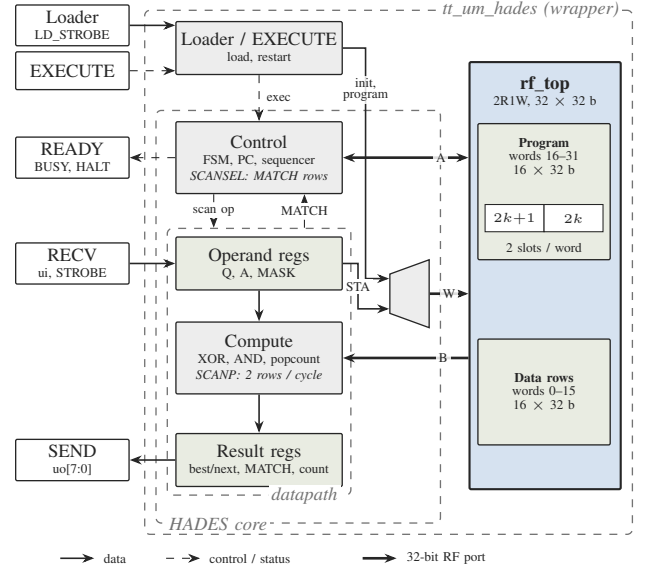

Figure~\ref{fig:arch-top} shows \hades{} at block level. The Tiny Tapeout top level wraps a core and a single register-file macro, \texttt{rf\_top}~\cite{munaut2025rftop}. The wrapper contains the program loader and the \isa{EXECUTE} edge detector. The core contains the control unit, the operand registers, the compute datapath, the result registers, a write arbiter, and the output multiplexer behind the \isa{SEND} pins. The core reaches memory only through the macro's ports.

\textbf{Design principle: one memory holds everything.} The macro is a $32\times32$-bit register file with two read ports and one write port (2R1W) and synchronous read. It serves three roles. It stores the dataset, it stores the program, and, through \isa{STA}, it provides scratch space for multi-pass programs. \hades{} therefore needs no separate instruction memory.

\hades{} divides the 32 words into $16$ \emph{data rows} (words $0$--$15$, $32$ bits each) and $16$ \emph{program words} (words $16$--$31$). Each program word packs two $16$-bit instructions, the even-numbered slot in the low half and the odd-numbered slot in the high half, giving a $32$-slot, $64$-byte program. Data and program have the same capacity, $64$ bytes each. This $1{:}1$ split is a \hades{} design choice enabled by the macro's flat $32$-word organisation. The macro does not distinguish data from program, and the \hades{} row-index and program-counter widths fix the partition. This paper does not explore other splits.

The two read ports and the write port are allocated so that instruction fetch and data access do not collide.
\begin{itemize}
\item Read port~B carries data-row reads.
\item Read port~A carries the instruction fetch, addressed by the next program counter so that the synchronous read returns the instruction word in time for the cycle in which it executes. It also carries the second row of a dual-row scan (Section~\ref{sec:scanext}).
\item The write port is shared by program loading, the post-reset initialisation of the program area, and the \isa{STA} instruction. A fixed priority of initialisation over loader over \isa{STA} resolves collisions in the write arbiter.
\end{itemize}

Data rows are retained through reset. The program is cleared, because a reset re-initialises the program area to \isa{HALT}.

\subsection{Architectural state and instruction set}

The programmer-visible state is small, and each register has a distinct job. Three $32$-bit working registers hold the query $Q$, an accumulator $A$, and a bitwise mask $M$. $M$ resets to all ones, and $Q$ and $A$ have no reset and are undefined until first written. Separating the three lets a program vary the query, the mask or the threshold per iteration while the scan datapath stays the same.

A set of result registers records the outcome of the most recent scan: a primary \emph{best} value and row index, a \emph{next} value and row index (valid after \isa{NEAREST2}), a $16$-bit \isa{MATCH} bitmap with one bit per data row, and a five-bit \isa{count}. Values are six bits wide so that a distance or popcount of $32$ is representable. The reset and \isa{INIT} value of the distance registers of the \isa{NEAREST} family is $\mathtt{0x3F}$, read as ``no row recorded'', whereas \isa{ARGMAX_POP} initialises its value to $0$. The chip reset alone sets the result registers to these identities. Two flags, \isa{MIN_VALID} and \isa{THRESHOLD_HIT}, and a five-bit program counter complete the state.

Instructions are $16$ bits wide with a four-bit opcode and three four-bit operand fields (Table~\ref{tab:isa}). Twelve opcodes are defined and four are reserved. The set divides into four groups.
\begin{itemize}
\item \emph{Register movement.} Fill $Q$, $A$ or $M$ from a data row (\isa{LDQ}, \isa{LDA}, \isa{LDM}) or from the host one byte at a time (\isa{RECV}), write $A$ back to a data row (\isa{STA}), and reset $M$ (\isa{UNMASK}).
\item \emph{Scan.} The scan family (\isa{SCAN}, \isa{SCANP}, \isa{SCANSEL}) is the only way to apply the compute datapath.
\item \emph{Output.} \isa{SEND} places a selected register on the output pins.
\item \emph{Control.} \isa{BRANCH} and \isa{HALT} provide control flow, with eight branch conditions drawn from the flags, the match and count registers, and whether $A$ is zero.
\end{itemize}

The only arithmetic in the design is inside the scans, and all other instructions move bytes or steer control. A reduction can produce an intermediate value in $A$, \isa{STA} can store it, and \isa{LDA}, \isa{LDQ} or \isa{LDM} can bring it back, which is how a multi-pass program uses the data rows as scratch space.

\begin{table}[t]
\centering
\caption{Instruction set (opcode in hexadecimal, bits 15--12). Fields A, B, C are the three low nibbles; $n$ is the number of rows scanned and $k$ the number of rows selected by \isa{MATCH}.}
\label{tab:isa}
\footnotesize
\setlength{\tabcolsep}{3pt}
\begin{tabular}{@{}ll P{2.9cm} l@{}}
\toprule
Opc. & Mnemonic & Meaning & Cycles \\
\midrule
\texttt{0x0} & \isa{LDQ} & $Q\leftarrow$ row & 2 \\
\texttt{0x1} & \isa{LDA} & $A\leftarrow$ row & 2 \\
\texttt{0x2} & \isa{LDM} & $M\leftarrow$ row & 2 \\
\texttt{0x3} & \isa{STA} & row $\leftarrow A$ & 1 \\
\texttt{0x4} & \isa{SCAN} & scan rows in order & $n{+}1$ \\
\texttt{0x5} & \isa{RECV} & host byte into a lane of $Q$/$A$/$M$ & host-gated \\
\texttt{0x6} & \isa{SEND} & drive output byte, pulse \isa{READY} & 1 \\
\texttt{0x7} & \isa{BRANCH} & signed relative branch & 1 \\
\texttt{0x8} & \isa{HALT} & stop, deassert \isa{BUSY} & 1 \\
\texttt{0x9} & \isa{UNMASK} & $M\leftarrow\mathtt{0xFFFFFFFF}$ & 1 \\
\texttt{0xA} & \isa{SCANP} & as \isa{SCAN}; two rows per cycle for \isa{EQUAL}, \isa{REDUCE_OR}, \isa{REDUCE_AND} & $1{+}\lceil n/2\rceil$ \\
\texttt{0xB} & \isa{SCANSEL} & as \isa{SCAN}, on rows set in \isa{MATCH} & $1{+}k$ \\
\texttt{0xC}--\texttt{0xF} & \multicolumn{3}{l}{reserved (behaviour undefined)} \\
\bottomrule
\end{tabular}
\end{table}

\subsection{Scan operations}

\textbf{Insight: the workloads differ in what a scan keeps, not in how it reads.} A scan visits data rows in ascending order, applies one of eight operations, and updates the result registers. A base row and a length give the row range, which is clamped at row $15$. A mode bit selects whether the accumulators are first set to their identity (\isa{INIT}) or carry over from a previous scan (\isa{ACC}), so chained scans can compose several row ranges into one result. The operations act on the masked distance $d_i$ or the masked population $p_i=\operatorname{pop}(D_i\wedge M)$ and fall into three groups.

\isa{NEAREST}, \isa{NEAREST2} and \isa{WITHIN} compute $d_i$ for every row. \isa{NEAREST} keeps the smallest distance and its row, with ties resolved to the lowest row index. \isa{NEAREST2} also keeps the second smallest, so that the two entries are the first two rows in order of $(d_i, i)$. \isa{WITHIN} behaves like \isa{NEAREST} and in addition records every row with $d_i\le A[7:0]$ in \isa{MATCH}, counts them, and sets \isa{THRESHOLD_HIT}.

\isa{EQUAL} records in \isa{MATCH} and \isa{count} the rows whose masked difference is zero.

\isa{REDUCE_OR}, \isa{REDUCE_AND} and \isa{REDUCE_POP} fold the masked rows into $A$ by bitwise OR, bitwise AND or summed popcounts, and \isa{ARGMAX_POP} keeps the row with the largest $p_i$. Because the reductions leave their result in $A$, \isa{STA} can store it for a later pass.

\subsection{Dual-row and selective scans}
\label{sec:scanext}

\isa{SCANP} and \isa{SCANSEL} target two limits of a one-row-per-cycle scan. Both leave the compute datapath unchanged, and programs that do not use them are unaffected.

\textbf{SCANP: use the idle read port.} The scan is limited by one row per cycle, yet the macro has a second read port that a plain \isa{SCAN} uses only for instruction fetch. The popcount tree is the expensive part of the datapath, so the operations that need no popcount can use the second port at little cost. \isa{SCANP} has the encoding and the architectural result of \isa{SCAN}, and only the cycle count differs. For the three operations that need no popcount, \isa{EQUAL}, \isa{REDUCE_OR} and \isa{REDUCE_AND}, \isa{SCANP} reads two rows per cycle, so $n$ rows take $1+\lceil n/2\rceil$ cycles, nine cycles for sixteen rows against seventeen for \isa{SCAN}. For every other operation \isa{SCANP} executes exactly as \isa{SCAN}, because the single popcount tree bounds those operations. No hazard is visible to the program.

\textbf{SCANSEL: make the match bitmap an operand.} A predicate scan already produces \isa{MATCH}. Without a way to consume it, a filter-then-refine query would have to leave the engine between its two stages. \isa{SCANSEL} applies the same operation and mode to the rows whose \isa{MATCH} bit is set when the instruction issues, in ascending order, skipping unselected rows. It has no row range, because the selection is the whole bitmap. A selection of $k$ rows takes $1+k$ cycles, and a full bitmap is equivalent to a sixteen-row \isa{SCAN}.

Because the selection is captured at issue, an \isa{INIT}-mode \isa{EQUAL} or \isa{WITHIN} \isa{SCANSEL} rebuilds \isa{MATCH} from the selected rows only, which narrows a candidate set in place. A filter-then-refine query therefore runs entirely inside the engine without a round trip through the host.

\subsection{Host interface and execution model}

The chip is driven through the eight input pins \texttt{ui[7:0]}, eight output pins \texttt{uo[7:0]} and, of the bidirectional pins, three used as inputs (\isa{STROBE}, \isa{LD_STROBE}, \isa{EXECUTE}) and three as outputs (\isa{READY}, \isa{BUSY}, \isa{HALT}).

After reset the engine spends $16$ clocks writing \isa{HALT} into all program slots while ignoring the loader. The host then loads the program through the byte loader. On each rising edge of \isa{LD_STROBE} the byte on \texttt{ui} is captured, two bytes form an instruction (high byte first), and instructions fill slots $0,1,2,\dots$. The loader writes the even slot of a program word at once, with \isa{HALT} in the odd half, and completes the word when the odd slot arrives. Slots that are never written remain \isa{HALT}. Reloading the program while it executes is not supported.

A rising edge on \isa{EXECUTE} starts the program at slot $0$ two clocks later. An \isa{EXECUTE} issued while the program is running or halted restarts it from slot $0$. A restart resets only the control unit (the state machine, program counter, scan sequencing and \isa{SCANSEL} selection state, and the operation-select flops). $Q$, $A$, $M$, the result registers, \isa{MATCH}, \isa{count}, the flags and the data rows keep their values, so a restart during a scan leaves partial results in the result registers. Only the chip reset clears $M$ and the result registers, and $Q$ and $A$ are never cleared. This lets a dataset in the data rows persist across queries and across program restarts.

At run time the host exchanges data with the program through \isa{RECV} and \isa{SEND}. \isa{RECV} stalls until the host raises \isa{STROBE} with a byte on \texttt{ui}, and writes the byte into one byte lane of $Q$, $A$ or $M$, so a $32$-bit operand takes four \isa{RECV} instructions. \isa{SEND} drives one byte onto \texttt{uo} and pulses \isa{READY} for one cycle. There is no hardware flow control: the host must consume each byte in that cycle and must consume all of an iteration's output before driving the next \isa{RECV}. The twelve \isa{SEND} sources are the best index and value, the four bytes of $A$, the two bytes of \isa{MATCH}, the count, the flags, and the next index and value.

\subsection{Microarchitecture and cycle model}

\begin{figure*}[t]
\centering
\begin{tikzpicture}[hades, x=1cm, y=1cm,
    core/.append style={font=\scriptsize}, reg/.append style={font=\scriptsize}, io/.append style={font=\scriptsize},
    mem/.append style={font=\scriptsize},
    rev2/.style={core, densely dashed, line width=0.8pt, fill=white},
    lab/.style={wirelabel, font=\fontsize{6}{6.6}\selectfont, inner sep=1pt}]
  \providecommand{\hsm}{\fontsize{6}{6.6}\selectfont}

  \node[io, minimum width=1.4cm, minimum height=0.5cm] (pLD) at (0.75,8.20) {Loader\\[-2pt]{\hsm ui, uio[4]}};
  \node[io, minimum width=1.4cm, minimum height=0.5cm] (pEX) at (0.75,7.40) {EXECUTE};
  \node[io, minimum width=1.4cm, minimum height=0.5cm] (pHS) at (0.75,6.10) {BUSY\\[-2pt]{\hsm HALT}};
  \node[io, minimum width=1.4cm, minimum height=0.5cm] (pSD) at (0.75,3.60) {SEND\\[-2pt]{\hsm uo, READY}};
  \node[io, minimum width=1.4cm, minimum height=0.5cm] (pRV) at (0.75,1.60) {RECV\\[-2pt]{\hsm ui, uio[3]}};

  \node[core, minimum width=2.5cm, minimum height=0.95cm] (ldr) at (3.200,7.80) {Loader / EXEC};

  \node[core, minimum width=1.7cm, minimum height=0.5cm] (ini) at (8.200,8.55) {Init: HALT fill};
  \node[core, minimum width=2.6cm, minimum height=0.70cm] (arb) at (13.250,7.75) {Write arbiter\\[-2pt]{\hsm init $>$ loader $>$ STA}};
  \node[core, minimum width=3.4cm, minimum height=1.75cm] (ctl) at (9.850,5.65) {};
  \node[anchor=north, align=center, font=\scriptsize] at ([yshift=-2pt]ctl.north) {\textbf{Control}\\[-2pt]{\hsm FSM, PC, scan}};
  \node[rev2, minimum width=2.7cm, minimum height=0.42cm, font=\hsm] (sel) at (9.850,5.07) {SCANSEL select};
  \node[core, minimum width=2.6cm, minimum height=0.70cm] (ins) at (13.250,5.75) {Instruction select\\[-2pt]{\hsm PC / run}};
  \node[core, minimum width=2.6cm, minimum height=1.00cm] (cmp) at (13.250,3.60) {Compute\\[-2pt]{\hsm XOR, AND, popcount}};
  \node[rev2, minimum width=2.6cm, minimum height=0.90cm] (sr) at (13.250,1.60) {SCANP second row\\[-2pt]{\hsm XOR, AND}};
  \node[reg,  minimum width=2.6cm, minimum height=1.10cm] (reg) at (9.850,3.60) {Operand registers\\[-2pt]{\hsm Q, A, MASK}};
  \node[reg,  minimum width=2.7cm, minimum height=1.30cm] (res) at (6.400,3.60) {Result registers\\[-2pt]{\hsm best/next, match, count}};
  \node[core, minimum width=2.6cm, minimum height=1.40cm] (omx) at (3.200,3.60) {Output mux};

  \node[mem, minimum width=2.0cm, minimum height=8.2cm, anchor=south west] (rf) at (15.700,0.60) {};
  \node[anchor=north, font=\scriptsize\bfseries, align=center] at ([yshift=-2pt]rf.north) {rf\_top\\[-2pt]{\hsm\mdseries 2R1W, 32 $\times$ 32 b}};
  \node[reg, minimum width=1.75cm, minimum height=2.95cm] (rfp) at (16.700,6.45) {};
  \node[font=\hsm, align=center, anchor=north] at ([yshift=-2pt]rfp.north) {\textbf{Program}\\[-1pt]words 16--31};
  \node[draw, line width=0.4pt, minimum width=0.80cm, minimum height=0.36cm, font=\hsm, inner sep=1pt, fill=white] at (16.300,5.62) {$2k{+}1$};
  \node[draw, line width=0.4pt, minimum width=0.80cm, minimum height=0.36cm, font=\hsm, inner sep=1pt, fill=white] at (17.100,5.62) {$2k$};
  \node[reg, minimum width=1.75cm, minimum height=3.0cm, font=\hsm, align=center] (rfd) at (16.700,2.55) {\textbf{Data rows}\\[-1pt]words 0--15};

  \begin{scope}[on background layer]
    \node[group, fit=(ini)(arb)(ctl)(ins)(cmp)(sr)(reg)(res)(omx), inner sep=4pt] (core) {};
    \node[group, fit=(ldr)(core)(rf), inner sep=3pt] (wrap) {};
  \end{scope}
  \node[grouplabel, anchor=west] at ([xshift=4pt]core.south west) {HADES core};
  \node[grouplabel, anchor=east] at ([xshift=-4pt]wrap.north east) {tt\_um\_hades (wrapper)};

  \newcommand{\tagnode}[3]{\node[tag] at (#1,#2) {#3};}
  \newcommand{\tagjoin}[2]{\fill (#1,#2) circle (0.04);}

  \draw[data] (pLD.east) -- ([yshift=0.40cm]ldr.west);
  \draw[ctl]  (pEX.east) -- ([yshift=-0.40cm]ldr.west);
  \draw[bus]  ([yshift=-0.10cm]ldr.east) -- ([yshift=-0.05cm]arb.west) node[lab, pos=0.5, above=1pt] {pm\_wr};
  \draw[bus]  (ini.east) -| ([xshift=-0.65cm]arb.north) node[lab, pos=0.25, above=1pt] {init};
  \draw[ctl]  (ldr.south) -- ++(0,-0.566) -- ([xshift=-1.15cm]ctl.north |- {0,6.75}) -- ([xshift=-1.15cm]ctl.north) node[lab, pos=0.25, above=1pt] {exec\_req};

  \draw[ctl]  ([yshift=0.45cm]ctl.west) -- (pHS.east) node[lab, pos=0.5, above=1pt] {BUSY, HALT};
  \draw[ctl]  ([yshift=0.10cm]ctl.west) -| (omx.north) node[lab, pos=0.30, above=1pt] {emit\_src};

  \draw[ctl]  ([xshift=-0.90cm]res.north) |- ([yshift=-0.25cm]ctl.west);
  \node[lab, fill opacity=1, inner sep=1.5pt] at (6.850,5.40) {MATCH[15:0], flags};   
  \draw[ctl]  ([yshift=-0.60cm]ctl.west) -| ([xshift=1.10cm]res.north);
  \node[lab, fill opacity=1, inner sep=1pt, align=center] at (7.825,5.05) {Scan\\[-2pt]control};   

  \draw[ctl]  (ctl.south) -- (reg.north) node[lab, pos=0.5] {write en.};
  \tagnode{8.800}{4.48}{b}   \draw[data] (8.800,4.31) -- ([xshift=-1.05cm]reg.north);
  \tagnode{11.100}{4.48}{a}  \draw[ctl]  (11.100,4.65) -- ([xshift=1.25cm]ctl.south);

  \draw[ctl, bus] ([yshift=0.65cm]ctl.east) -- (rf.west |- {0,6.30}) node[lab, pos=0.58, above=1pt] {A addr, SCANP row};
  \draw[bus]  (rf.west |- {0,5.75}) -- (ins.east);                   
  \tagjoin{15.100}{5.75}  \draw (15.100,5.75) -- (15.100,5.47);  \tagnode{15.100}{5.30}{e}
  \draw[data] (ins.west) -- ([yshift=0.10cm]ctl.east);                  
  \node[lab] at (11.750,5.98) {instr};
  \draw[ctl]  ([yshift=-0.45cm]ctl.east) -- (11.820,5.20);  \tagnode{11.990}{5.20}{f}
  \tagnode{12.150}{4.45}{f}  \draw[ctl] (12.150,4.28) -- ([xshift=-1.10cm]cmp.north);
  \draw[ctl, bus] ([yshift=-0.75cm]ctl.east) -- (15.300,4.90) -- (15.300,3.95) -- (rf.west |- {0,3.95});
  \node[lab, fill opacity=1, inner sep=1.5pt] at (13.000,4.90) {B addr, SCANSEL row};   
  \draw[bus]  (rf.west |- {0,3.55}) -- ([yshift=-0.05cm]cmp.east);       
  \tagjoin{15.100}{3.55}  \draw (15.100,3.55) -- (15.100,3.27);  \tagnode{15.100}{3.10}{b}

  \node[lab, anchor=east] at (15.65,6.03) {port A};
  \node[lab, anchor=east] at (15.65,3.75) {port B};

  \draw[bus]  (arb.east) -- (rf.west |- {0,7.75}) node[lab, pos=0.5, above=1pt] {W};
  \tagnode{12.600}{7.05}{a}  \draw[data] (12.600,7.22) -- ([xshift=-0.65cm]arb.south);

  \draw[data] ([yshift=0.15cm]reg.east) -- ([yshift=0.15cm]cmp.west);
  \draw[data] ([yshift=-0.30cm]reg.east) -- (11.550,3.30) -- (11.550,1.60) -- (sr.west);
  \draw[data] ([yshift=0.30cm]reg.west) -- ([yshift=0.30cm]res.east);   \node[lab] at (8.150,4.07) {A};
  \draw[data] ([yshift=-0.25cm]res.east) -- ([yshift=-0.25cm]reg.west);   \node[lab] at (8.150,3.18) {A$\leftarrow$};
  \draw ([xshift=-0.65cm]reg.south) -- (9.200,2.87);  \tagnode{9.200}{2.70}{a}
  \draw[data] (pRV.east) -| ([xshift=0.75cm]reg.south) node[lab, pos=0.12, above=1pt] {RECV byte};
  \draw ([xshift=-0.45cm]cmp.south) -- (12.800,2.87);  \tagnode{12.800}{2.70}{c}
  \draw[data] (5.800,2.58) -- ([xshift=-0.60cm]res.south);  \tagnode{5.800}{2.41}{c}
  \draw (sr.south) -- (13.250,0.97);  \tagnode{13.250}{0.80}{d}
  \draw[data] (6.700,2.58) -- ([xshift=0.30cm]res.south);  \tagnode{6.700}{2.41}{d}
  \draw[data] (13.900,2.28) -- ([xshift=0.65cm]sr.north);  \tagnode{13.900}{2.45}{e}

  \draw[data] (res.west) -- (omx.east);
  \draw[data] (omx.west) -- (pSD.east);
  \draw[data] (3.200,2.58) -- (omx.south);  \tagnode{3.200}{2.41}{a}

  \draw[data] (0.60,-0.55) -- ++(0.45,0) node[right, font=\hsm] {data};
  \draw[ctl]  (2.10,-0.55) -- ++(0.45,0) node[right, font=\hsm] {control / status};
  \draw[bus]  (4.60,-0.55) -- ++(0.45,0) node[right, font=\hsm] {wide bus / RF port};
  \node[rev2, minimum width=0.45cm, minimum height=0.22cm] at (8.30,-0.55) {};
  \node[anchor=west, font=\hsm] at (8.60,-0.55) {SCANP / SCANSEL path};
  \node[tag, inner sep=0.5pt] at (13.40,-0.55) {x};
  \node[anchor=west, font=\hsm] at (13.65,-0.55) {same letter = same net};
\end{tikzpicture}
\caption{Detailed \hades{} microarchitecture. Circled letters mark the same net at both ends (a: accumulator, b: port-B data, c: compute result, d: second-row result, e: port-A data, f: operation select). Dashed outlines mark the \isa{SCANP} second-row path and the \isa{SCANSEL} select logic. Clock and reset are global and not drawn.}
\label{fig:arch-micro}
\end{figure*}
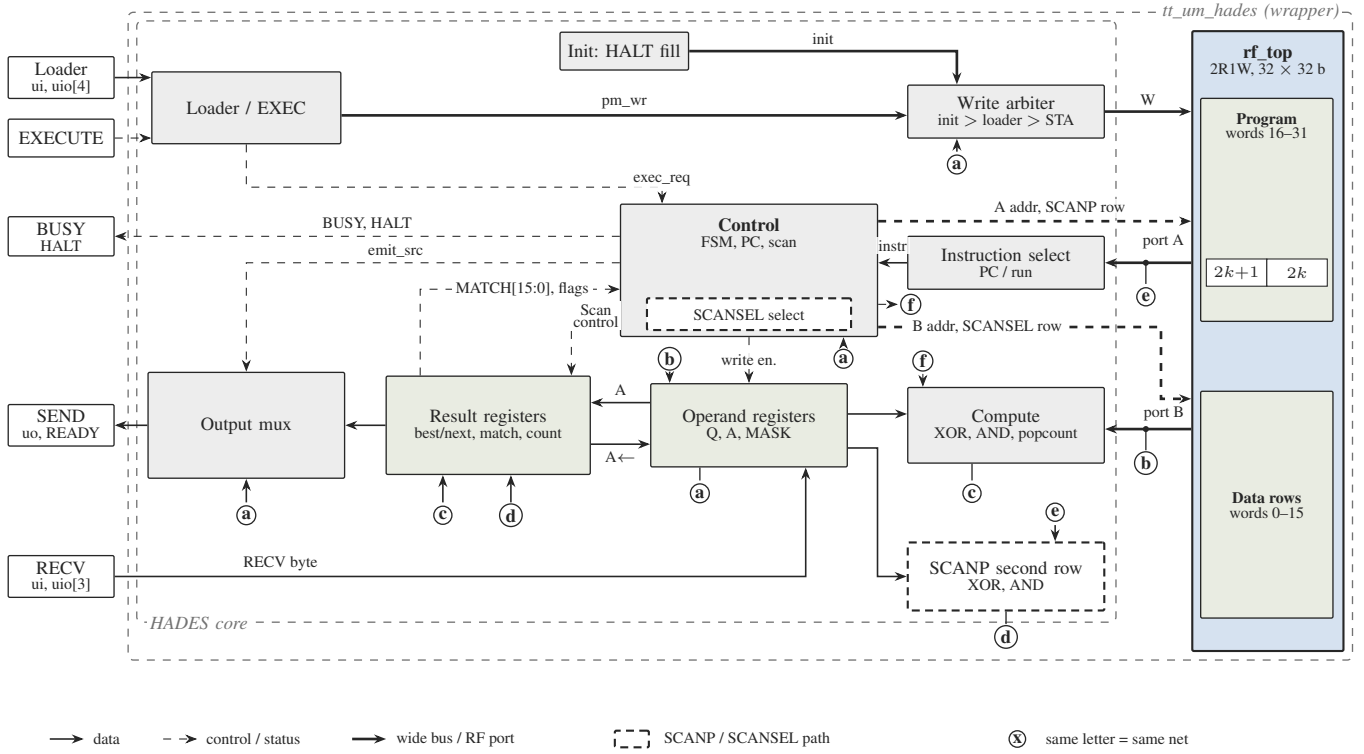

Figure~\ref{fig:arch-micro} refines Figure~\ref{fig:arch-top} to the level of the control path and datapath. It is a block-level view of the routing and does not reproduce the RTL. Control is a five-state finite-state machine (fetch, scan, wait-for-byte, load-wait, halt) together with the program counter, a scan-row counter and, for \isa{SCANSEL}, a sixteen-bit selection register and the lowest-set-bit logic. An instruction-select stage chooses between the fetched instruction and \isa{HALT}, so the core is idle until the first \isa{EXECUTE}. Scan control drives the operation, the row sequence and the ``reset-to-identity'' signal to the operand, compute and result blocks. Dashed blocks mark the \isa{SCANP} and \isa{SCANSEL} logic.

The operand registers hold $Q$, $A$ and $M$, each updated by byte lane from the host or in full from a row read. The compute block receives a row from port~B and forms, byte lane by byte lane, either $(D_i\oplus Q)\wedge M$ or $D_i\wedge M$ depending on the operation. A $32$-input popcount and a zero detect follow. The \isa{SCANP} path reads port~A through a second masked XOR and AND network and a zero detector, with no second popcount. Result registers hold the best and next entries with their comparators, the match bitmap and count, and the flags, and the output multiplexer selects among them for \isa{SEND}.

Instruction fetch addresses the macro with the next program counter, and the low bit of the program counter selects the correct half of the returned word. The same synchronous read determines the timing of the register-load instructions. A load issues its read in the fetch cycle and writes the destination register in the following cycle, so \isa{LDQ}, \isa{LDA} and \isa{LDM} take two cycles and the next instruction sees the new value. A \isa{STA} followed by a load of the same row therefore moves $A$ into $Q$ or $M$ on chip. Scans overlap the read of the next row with the compute on the current one.

The resulting cycle model (Table~\ref{tab:isa}) is: $2$ cycles for each register load; $n+1$ for a scan of $n$ rows ($17$ for a full sixteen-row scan); $1+\lceil n/2\rceil$ for a dual-row \isa{SCANP} ($9$ for sixteen rows); $1+k$ for \isa{SCANSEL} with $k$ selected rows; and one cycle for every other instruction except \isa{RECV}, which is gated by the host. Timing is independent of data except through \isa{SCANSEL}'s $k$ and through \isa{BRANCH}. These formulas specify the design, and Section~\ref{sec:eval} checks them against RTL and gate-level simulation.

\section{HADES Functionality and Workloads}
\label{sec:workloads}

\subsection{Workload semantics}

We define ten workloads by pure-function semantics over a flat dataset of $N$ rows $D[0..N{-}1]$, a $32$-bit query $Q$, a mask $M$ and an eight-bit threshold $T$, with $d_i=\operatorname{pop}((D[i]\oplus Q)\wedge M)$ and $p_i=\operatorname{pop}(D[i]\wedge M)$. Index ties always break toward the lowest index. The definitions are independent of \hades{}: the reference code that runs on every evaluation platform implements the same function, and the output is a canonical byte string against which every platform is checked byte for byte. Table~\ref{tab:workloads} summarises them.

\textbf{Why the ISA fits.} Each workload reduces to the same shape: bring the operands in, run one scan whose operation selects what is kept, and send a short record. The workloads differ in the scan operation, in the operand that varies per query, and in the \isa{SEND} sequence. This is why a small instruction set suffices.

The \emph{nearest} workload (\isa{NEAREST}) returns the lowest row index at the minimum distance and that distance. \emph{Nearest-two} (\isa{NEAREST2}) returns the first two rows in $(d_i,i)$ order together with their distances. \emph{Equality} (\isa{EQUAL}) returns the bitmap of rows whose masked bits equal the query and their count. \emph{Within-threshold} (\isa{WITHIN}) returns the nearest row \emph{and} the bitmap and count of rows with $d_i\le T$ and the \isa{THRESHOLD_HIT} flag. \emph{Masked similarity} is \isa{NEAREST} with a non-trivial mask, which selects a subset of the $32$ dimensions. By construction it exercises the mask without new control flow.

Three workloads run over the stored rows. \isa{REDUCE_OR} and \isa{REDUCE_AND} return the bitwise OR and the bitwise AND of the masked rows, and \isa{ARGMAX_POP} returns the row with the largest masked population and that population, with $(0,0)$ if all populations are zero.

Two further entries are not new semantics of the same kind. The \emph{dual-row} entry denotes executing the equality, OR or AND workloads with \isa{SCANP}. Its result is by definition that of \isa{SCAN}, so it compares cycle counts \emph{within} \hades{} and adds no new function. The \emph{filter-and-refine} workload uses \isa{SCANSEL}. A first stage of \isa{EQUAL} over all rows, parameterised by a key and a mask, selects a candidate bitmap of $k$ rows, and a second stage runs \isa{NEAREST} with a new query over the selected rows only. It returns the index and distance of the nearest selected row and the selection size $k$. Filter-and-refine is the one workload that composes two scans through \isa{MATCH}, and it is the reason \isa{SCANSEL} exists.

\begin{table*}[t]
\centering
\caption{Workloads and their result records. The \isa{SEND} sequence is the canonical record; ``slots'' is the number of instruction slots of the resident-query loop template (Section~\ref{sec:templates}), counted from the program templates. The loop form ends in a \isa{BRANCH} and the restart form in a \isa{HALT}, with the same slot count; the filter-and-refine count of $19$ applies to both.}
\label{tab:workloads}
\footnotesize
\setlength{\tabcolsep}{4pt}
\begin{tabular}{@{}lP{2.9cm}P{4.6cm}P{6.0cm}rr@{}}
\toprule
Workload & Scan operation & Result & \isa{SEND} sequence & Bytes & Slots \\
\midrule
Nearest & \isa{NEAREST} & lowest index of minimum distance, distance & \isa{BEST_IDX}, \isa{BEST_VAL} & 2 & 8 \\
Nearest-two & \isa{NEAREST2} & two nearest entries by $(d_i,i)$ & \isa{BEST_IDX}, \isa{BEST_VAL}, \isa{NEXT_IDX}, \isa{NEXT_VAL} & 4 & 10 \\
Equality & \isa{EQUAL} & match bitmap, count & \isa{MATCH_LO}, \isa{MATCH_HI}, \isa{COUNT} & 3 & 9 \\
Within-threshold & \isa{WITHIN} & nearest, bitmap of $d_i\le T$, count, flags & \isa{BEST_IDX}, \isa{BEST_VAL}, \isa{MATCH_LO}, \isa{MATCH_HI}, \isa{COUNT}, \isa{FLAGS} & 6 & 13 \\
Masked similarity & \isa{NEAREST}, masked & as nearest with non-trivial mask & as nearest & 2 & 8 (+4) \\
OR reduction & \isa{REDUCE_OR} & OR of masked rows & \isa{A_BYTE0}--\isa{A_BYTE3} & 4 & 10 \\
AND reduction & \isa{REDUCE_AND} & AND of masked rows & \isa{A_BYTE0}--\isa{A_BYTE3} & 4 & 10 \\
Argmax-popcount & \isa{ARGMAX_POP} & row of maximum masked population, value & \isa{BEST_IDX}, \isa{BEST_VAL} & 2 & 8 \\
Dual-row & \isa{SCANP} & as equality, OR, AND & as equality, OR, AND & 3--4 & 9--10 \\
Filter-and-refine & \isa{EQUAL} then \isa{SCANSEL} \isa{NEAREST} & nearest among selected rows, $k$ & \isa{BEST_IDX}, \isa{BEST_VAL}, \isa{COUNT} & 3 & 19 \\
\bottomrule
\end{tabular}
\end{table*}

\subsection{Program templates and result records}
\label{sec:templates}

Each workload is expressed as a short program that fits in the $32$ slots. \isa{LOADQ} and \isa{LOADMASK} are assembler pseudo-instructions (not ISA opcodes): each expands to four \isa{RECV} instructions that fill the four byte lanes of $Q$ or $M$, and so occupies four program slots. The resident-query form of the nearest workload loads a query, scans all rows, emits the record, and loops, with the dataset remaining in the data rows between iterations:
\begin{small}
\begin{verbatim}
loop:  LOADQ
       SCAN 0, 16, NEAREST, INIT
       SEND BEST_IDX
       SEND BEST_VAL
       BRANCH ALWAYS, loop
\end{verbatim}
\end{small}
The equality and reduction workloads differ only in the scan operation, the operand that varies per query (the query for equality, the mask for the population workloads), and the \isa{SEND} sequence. Within-threshold additionally loads the threshold into $A[7:0]$ with a single \isa{RECV}.

The two-stage workload chains two scans through \isa{MATCH}:
\begin{small}
\begin{verbatim}
       LOADQ              ; selection key
       LOADMASK           ; bench: ~Sel&0xFFFF
       SCANP 0, 16, EQUAL, INIT
       UNMASK
       LOADQ              ; refine query
       SCANSEL NEAREST, INIT
       SEND BEST_IDX
       SEND BEST_VAL
       SEND COUNT
       HALT
\end{verbatim}
\end{small}
\isa{UNMASK} restores a full mask for the refinement, and the first-stage scan may equally be a plain \isa{SCAN}. In the evaluation the first-stage key and mask ($\sim\!\mathit{Sel}\,\&\,\mathtt{0xFFFF}$, over a dataset built for the purpose) make \isa{EQUAL} select a chosen subset of $k$ rows. This benchmark construction gives a controlled selection size. \isa{SCANSEL} consumes whatever \isa{MATCH} holds, however it was produced, so the construction imposes no architectural requirement.

The evaluation uses two forms of each template: one ends in \isa{BRANCH} back to the start, and one ends in \isa{HALT} followed by a host \isa{EXECUTE}. They differ only in who restarts the program and occupy the same number of slots. The template above is the \isa{HALT} form.

The program also loads the dataset. Because a $32$-bit row costs four \isa{RECV} instructions and one \isa{STA}, a program holds at most six rows of data in addition to a closing \isa{HALT}. A sixteen-row dataset is therefore loaded by three successive programs, each preceded by a reset and followed by an \isa{EXECUTE}. Data rows survive these resets, so the dataset is resident after the third program.

The host tiles datasets larger than sixteen rows: it loads each sixteen-row tile in turn and merges per-tile records. The merge rules (lowest global index for the best-value operations, concatenated bitmaps and summed counts for the predicate operations, folded accumulators for the reductions) preserve the flat-dataset semantics. Section~\ref{sec:eval} returns to the cost of ingestion and tiling.

Result records are short and fixed in length for a given workload, between two and six bytes in Table~\ref{tab:workloads}. For the \isa{NEAREST} family, a distance register that no row has updated reads as $\mathtt{0x3F}$. A \isa{SCAN} always visits at least one row, so this arises only after a \isa{SCANSEL} with \isa{MATCH}${}=0$, the one case in which a scan reads no rows. \isa{ARGMAX_POP} instead initialises its value and index to $0$, so an empty selection yields $(0,0)$. The canonical \isa{SEND} order is part of the definition of each workload and is the same on every platform.

\subsection{Expressiveness: program-slot utilisation}

Program depth is the resource the research question names, and Table~\ref{tab:workloads} reports how much of it each template uses. The resident-query loops of the eight scan workloads use between $8$ and $13$ slots, and masked similarity adds a four-slot prologue when the mask is loaded once outside the loop. The longest template is filter-and-refine with $19$ slots. The RTL regression also runs seven application programs of $8$, $9$, $11$, $14$, $12$, $13$ and $16$ instruction slots, excluding the terminating \isa{HALT}. The longest, a four-pass population analytics program, uses the reductions, \isa{STA} and \isa{LDA}, takes no host input, and uses the data rows as scratch space. Every template and regression program fits within the $32$ slots. Two of them, filter-and-refine and the analytics program (which needs $17$ slots with its \isa{HALT}), would not fit in $16$.

The largest template uses roughly sixty percent of the available depth. The counts do not establish that $32$ is the minimum: no workload in the set requires a number of slots between $19$ and $32$, and a different workload set could change the answer.

\textbf{I/O dominates the templates.} Slot counts also depend on how operands are supplied. Because every operand arrives through four byte-wide \isa{RECV} instructions, I/O instructions dominate the templates (eleven of the $13$ slots of within-threshold, for example, are \isa{RECV} or \isa{SEND}). This is a direct consequence of the narrow host interface.

\section{Implementation}
\label{sec:impl}

\subsection{Design flow}

The \hades{} RTL is written in Veryl~\cite{veryl2026} (version 0.21.0), which generates the SystemVerilog used by simulation and physical design. The design comprises the wrapper, the core and its control, register, compute and result-register blocks. The operation-select flops of the compute datapath are replicated to limit fan-out on a timing-critical net, and the second-row consumer of \isa{SCANP} has its own copies. The register file is the \texttt{rf\_top} macro from the Tiny Tapeout community~\cite{munaut2025rftop}, instantiated unmodified. It is a hard macro in the layout, and the logic around it is mapped to SKY130 standard cells.

Physical design uses the Tiny Tapeout flow~\cite{tinytapeout} on a $2\times2$ tile (die $334.88\times225.76$\,\textmu m), with a $25$\,ns clock period ($40$\,MHz) and the macro placed at a fixed position and orientation. All results below are post-layout results from this flow.

\subsection{Physical results on a \texorpdfstring{$2\times2$}{2x2} tile}

Table~\ref{tab:phys} summarises the result. The design closes timing at $40$\,MHz with no setup or hold violation in any of the nine analysed corners. The worst setup slack, $+3.59$\,ns, is at the slow corner (\texttt{max\_ss}), and the worst hold slack is $+0.115$\,ns. The router finishes with zero design-rule violations, and layout-versus-schematic, antenna and power-grid checks pass.

The \texttt{rf\_top} macro occupies $15{,}744$\,\textmu m$^2$ and the standard cells (including tap, buffer and antenna cells) occupy $26{,}968$\,\textmu m$^2$, so the standard cells and the macro together occupy $56.5\%$ of the die. Relative to the whole core, $41.1\%$ ($29{,}851.8$\,\textmu m$^2$) is free of cells (excluding fill). Relative to the macro-free core ($56{,}820.2$\,\textmu m$^2$), the same free area is about $52.5\%$.

The flow's whole-chip power figure is a default-activity estimate and is not reported. Section~\ref{sec:eval} derives the energy estimates separately.

\begin{table}[t]
\centering
\caption{Hardening result on a $2\times2$ tile (post-layout). Percentages are computed from the reported areas, with fill cells excluded.}
\label{tab:phys}
\footnotesize
\setlength{\tabcolsep}{3pt}
\begin{tabular}{@{}lP{4.3cm}@{}}
\toprule
Item & Result \\
\midrule
Die / core area & $75{,}602.5$ / $72{,}564.6$ \textmu m$^2$ \\
\texttt{rf\_top} macro & $15{,}744.4$ \textmu m$^2$ \\
Standard cells & $26{,}968.4$ \textmu m$^2$ ($3{,}194$ cells) \\
\hspace{1em}multi-input combinational & $15{,}203.3$ \textmu m$^2$ \\
\hspace{1em}sequential & $5{,}669.2$ \textmu m$^2$ \\
\hspace{1em}timing-repair buffers & $3{,}996.3$ \textmu m$^2$ ($423$ cells) \\
\hspace{1em}tap cells & $970.9$ \textmu m$^2$ \\
\hspace{1em}clock tree, antenna, other & $1{,}128.7$ \textmu m$^2$ ($152$ cells) \\
Occupied by std cells + macro & $42{,}712.8$ \textmu m$^2$: $56.5\%$ of die \\
Std-cell util., macro-free core & $47.5\%$ \\
Free area (excl.\ fill) & $29{,}851.8$ \textmu m$^2$: $41.1\%$ of core, $52.5\%$ of macro-free core \\
Route wirelength / vias / nets & $112{,}779$ \textmu m / $20{,}322$ / $2{,}425$ \\
Global-route usage, met1/2/3/4 & $57.7$ / $53.2$ / $29.7$ / $15.7\%$, overflow $0$ \\
DRC (route, Magic) / LVS / antenna & $0$ / $0$ / $0$ \\
Worst setup (\texttt{max\_ss}) & $+3.59$\,ns ($+4.11$\,ns at \texttt{nom\_ss}) \\
Worst hold (\texttt{min\_ff}) & $+0.115$\,ns \\
Violations, 9 corners (setup/hold) & $0$ / $0$ \\
Worst clock skew & $0.46$--$0.79$\,ns \\
\bottomrule
\end{tabular}
\end{table}

\subsection{Cost of \texttt{SCANP} and \texttt{SCANSEL}}

To quantify the cost of the two scan variants, Table~\ref{tab:cost} compares the final design with the same design without \isa{SCANP} and \isa{SCANSEL} (the ten base opcodes), hardened on the same tile, macro placement and clock. The two instructions cost $+3{,}988$\,\textmu m$^2$ of standard-cell area ($+17.4\%$, $+390$ cells), raise detail-route wirelength by $25\%$, and the worst setup slack is $1.29$\,ns lower ($+4.88$ against $+3.59$\,ns). Slack remains positive in both. Each design was hardened once, so the slack difference includes whatever variation synthesis and placement introduce for a netlist change of this size; we did not repeat the runs, and the difference should be read as an indication of the cost and not as a causal estimate.

Global-route usage on the two lowest metal layers rises from about $44$--$47\%$ to $53$--$58\%$ of the (already derated) capacity, still without overflow. The two instructions were hardened together, so these figures cannot separate the cost of \isa{SCANP} from that of \isa{SCANSEL}. A separate hardening with \isa{SCANP} alone gave $+2.4$--$2.6$\,k\textmu m$^2$ of standard-cell area.

\begin{table}[t]
\centering
\caption{Final design and the base-opcode design without \isa{SCANP} and \isa{SCANSEL}, on the same $2\times2$ tile, macro placement and $25$\,ns clock (post-layout).}
\label{tab:cost}
\footnotesize
\setlength{\tabcolsep}{3pt}
\begin{tabular}{@{}lrrr@{}}
\toprule
Metric & Base ISA & Final & Change \\
\midrule
Std-cell area (\textmu m$^2$) & $22{,}980.8$ & $26{,}968.4$ & $+17.4\%$ \\
Std cells & $2{,}804$ & $3{,}194$ & $+390$ \\
Occupied, share of die & $51.2\%$ & $56.5\%$ & $+5.3$ pt \\
Route wirelength (\textmu m) & $90{,}005$ & $112{,}779$ & $+25.3\%$ \\
Vias & $16{,}873$ & $20{,}322$ & $+20.4\%$ \\
Global-route met1 / met2 & $47.1$ / $44.2\%$ & $57.7$ / $53.2\%$ & \\
Worst setup (\texttt{max\_ss}, ns) & $+4.88$ & $+3.59$ & $-1.29$ \\
\bottomrule
\end{tabular}
\end{table}

\subsection{Verification}

Verification has four layers.
\begin{enumerate}
\item[(i)] A Python assembler and cycle-accurate instruction-set model, with its own test suite ($102$ tests, all passing).
\item[(ii)] A self-checking Verilog testbench of the design ($84$ tests, $379$ checks, all passing).
\item[(iii)] A pin-level cocotb suite of $42$ tests that drives only the chip's pins (reset, loader, \isa{EXECUTE}, \isa{RECV} strobes, and the \isa{SEND} bytes), so the same tests apply to RTL and to a gate-level netlist. It covers the loader and restart behaviour, every instruction and branch condition, all scan operations and their \isa{INIT}/\isa{ACC} chains, \isa{SCANP} for every base and length including its fallback operations, \isa{SCANSEL} for empty, single and multiple selections, and fifteen golden workloads with checked cycle counts and result bytes. All $42$ pass at RTL.
\item[(iv)] The evaluation vectors of Section~\ref{sec:eval}, whose expected results come from a flat reference implementation and are cross-checked against the Python model on every query and against the RTL byte for byte.
\end{enumerate}
Gate-level confirmation on the post-layout netlist covers a subset of these and is described in Section~\ref{sec:gl}.

\section{Evaluation}
\label{sec:eval}

The evaluation asks five questions. Does the compact programmable engine accelerate the kernels? Does the advantage survive the cost of query transfer and of loading the dataset and program? How does the advantage change with batch size? What does the physical implementation cost? What does the measured microcontroller baseline tell us about power versus execution time? It does not assume that \hades{} is faster.

All \hades{} figures are cycle counts from RTL or gate-level simulation of the final design, converted at the specified $40$\,MHz clock ($1$ cycle $=25$\,ns), with an ideal host that holds \isa{STROBE} when \isa{RECV} needs it and consumes each \isa{SEND} byte immediately. Every number below carries one of the labels in Table~\ref{tab:labels}, and numbers taken at different boundaries are never combined.

\begin{table}[t]
\centering
\caption{Evidence labels used in this section.}
\label{tab:labels}
\footnotesize
\setlength{\tabcolsep}{3pt}
\begin{tabular}{@{}lP{6.4cm}@{}}
\toprule
Label & Meaning \\
\midrule
RTL & cycles from the cocotb/Icarus simulation of the final RTL; result bytes checked; ideal host \\
GL & the same runs on the post-layout gate-level netlist \\
FORMULA & closed-form cycle count, verified equal to RTL \\
MEAS & measured on the baseline boards after the correctness gate \\
MEAS-E & measured board power and energy per query of the baselines, with an inline shunt monitor \\
DERIVED & arithmetic on tabulated results (tiles $\times$ cycles, cycles $/f$, cycles $\times$ energy per cycle) \\
MODEL & measured byte counts with an assumed link rate \\
EST-M, EST-W & pre-silicon energy estimates (macro only; whole chip) \\
\bottomrule
\end{tabular}
\end{table}

\subsection{Methodology}

\paragraph{Platforms.} Table~\ref{tab:platforms} lists the platforms. E, the ESP8266, is the primary baseline and runs every set. E32, an ESP32, runs the same sets, vectors and kernels at three times the clock. N, an Arduino Nano, runs only the subset that fits its $2$\,KB of RAM and $32$\,KB of flash: the compute-only set, scaling up to $N=64$, the resident-query sets up to $B=16$ and the mask-density and per-query-mask sets at $B=256$. The selected-scan, filter-and-refine and end-to-end sets do not build for the Nano, so the Nano has no result for them. All three boards also appear in the energy measurement of Section~\ref{sec:energy}.

The baselines run plain C written from the workload semantics of Section~\ref{sec:workloads}, with one portable kernel per workload and no hand-written assembly, SIMD or data-layout tuning. On all three boards the compiler implements popcount as a call to a library routine. The ESP8266 core has no popcount instruction. Optimised baselines would be a separate experiment and are not included.

\begin{table}[t]
\centering
\caption{Platforms.}
\label{tab:platforms}
\footnotesize
\setlength{\tabcolsep}{3pt}
\begin{tabular}{@{}lP{6.3cm}@{}}
\toprule
ID & Configuration \\
\midrule
H & \hades{}, $2\times2$ tile, specified $40$\,MHz; RTL and gate-level simulation; ideal host; pre-silicon \\
E & ESP8266EX, Arduino core 3.1.2, $80$\,MHz, \texttt{-O2}, Wi-Fi off, cycle counter \texttt{CCOUNT}; all sets \\
N & Arduino Nano, ATmega328P, avr-gcc 7.3, $16$\,MHz, \texttt{-O2}, $16$-bit timer; restricted subset \\
E32 & ESP32-D0WD, Arduino-ESP32 core 3.3.12, $240$\,MHz, \texttt{-O2}, radio not started, cycle counter, interrupts masked in timed regions; all sets \\
\bottomrule
\end{tabular}
\end{table}

\paragraph{Boundaries.} Three measurement boundaries separate what each number includes.
\begin{itemize}
\item \emph{Compute-only} (B-C) is the scan kernel alone, with rows, query, mask and threshold already in working storage. On \hades{} it is the cycle count of the scan instruction.
\item \emph{Resident-query} (B-R) is one query against a resident dataset and a resident program: query ingress, the scan and result egress at byte level, without dataset or program load. On the microcontrollers this boundary also contains serialising each record and folding it into a running FNV-1a-64 hash (the on-device correctness gate), work that \hades{} does not perform; Section~\ref{sec:br} bounds its effect.
\item \emph{End-to-end} (B-E) adds the dataset ingestion and program loading needed to go from byte streams to result bytes, for a batch of $B$ queries over $N$ rows.
\end{itemize}
B-C and B-R exist on \hades{} at $N=16$ only, which is its data capacity. The host tiles larger datasets, and the host-side merge is excluded from chip time. End-to-end uses byte buffers in memory on E and E32 and models the physical link for \hades{} (Section~\ref{sec:be}).

\paragraph{Workloads and data.} The workloads are the ten of Section~\ref{sec:workloads}. We generate rows and queries from seeds in three distributions: uniform (D1), queries planted at known flip distances from stored rows (D2), and tie-heavy rows drawn from a pool of four values (D3, used for correctness and tie semantics). Dataset sizes are $N\in\{16,32,64,256,1024\}$ ($4096$ on E and E32 for context). Mask density is $b\in\{32,24,16,8,1\}$ bits, batch size is $B\in\{1,4,16,64,256,1024\}$ for repeated queries, and the \isa{SCANSEL} selection count is $k\in\{0,1,2,4,8,16\}$.

\paragraph{Correctness gate.} One set of vectors, with its expected result for every query, serves all platforms. Expected results come from a flat reference implementation and are cross-checked before any timing: against the Python instruction-set model on every query of every set, against the C reference compiled on an x86 host, and against the RTL for the $N=16$ sets. The RTL cross-checks are byte-exact and cover $1{,}675$ canonical cases, $1{,}920$ filter-and-refine cases in each of two variants, the $255$ compute-only cases, the resident-query sets ($70$ cells of $1{,}024$ queries), $300$ end-to-end runs, and $280$ tile-merge runs on the D2 and D3 distributions. A cell with any mismatch has its timing discarded. Timing on E, N and E32 uses $R=30$ repetitions after warm-up. Each timed region lasts at least $1$\,ms, and the timer overhead is calibrated and removed. The resident-query and end-to-end per-query times on these boards include serialising each record and folding it into a running hash, which replaces an output buffer, and no performance counters are available on the boards.

\paragraph{Presentation rules.} Absolute time per query is the primary cross-platform quantity. Cycles of \hades{} are never compared with cycles or times of a microcontroller; the only cycle comparison is between the two ESP baselines (Section~\ref{sec:bc}). Ratios appear only between numbers of the same boundary, and the differing scopes of the platforms apply to every ratio. The baseline results use two kernel engines: a constant-$N$ engine (compute-only, $N=16$) and a runtime-$N$ engine (all other sets). We compare only results from the same engine, and we report ratios against each baseline separately and never pool them across baselines.

\subsection{Cycle model verification}

Does the implementation obey the cycle model? Table~\ref{tab:isa} specifies the cycle model of Section~\ref{sec:arch}. We check it at instruction level with $154$ cases covering \isa{SCAN}, \isa{SCANP} and \isa{SCANSEL} over row counts $n\in\{1,2,3,15,16\}$, selection sizes $k\in\{0,1,2,4,8,16\}$ and all operations, together with \isa{LDx}, \isa{STA}, \isa{RECV}, \isa{SEND} and \isa{BRANCH}. In all $154$ cases the closed-form cycle count, the Python model and the RTL agree exactly [RTL, FORMULA], and the gate-level run agrees as well (Section~\ref{sec:gl}). A sixteen-row \isa{SCAN} takes $17$ cycles, a sixteen-row dual-row \isa{SCANP} takes $9$, and a \isa{SCANSEL} of $k$ rows takes $1+k$.

\subsection{Compute-only latency}\label{sec:bc}

Does the compact engine accelerate the kernels themselves? Table~\ref{tab:bc} gives the compute-only latency of the eight scan workloads at $N=16$. On \hades{} every one of the eight takes $17$ cycles, $425$\,ns, because the scan visits sixteen rows through the same single popcount tree regardless of the operation [RTL, GL].

\begin{table}[t]
\centering
\caption{Compute-only (B-C) latency, $N=16$, uniform data. H: RTL, $17$ cycles at $40$\,MHz. E, N, E32: measured, per kernel call, constant-$N$ engine, median over the cases of each workload; all three run the same $255$ cases.}
\label{tab:bc}
\footnotesize
\setlength{\tabcolsep}{4pt}
\begin{tabular}{@{}lrrrr@{}}
\toprule
Workload & H (ns) & E (ns) & N (ns) & E32 (ns) \\
\midrule
\isa{NEAREST} & 425 & \num{12809} & \num{171711} & \num{5192} \\
\isa{NEAREST2} & 425 & \num{13834} & \num{176211} & \num{5348} \\
\isa{EQUAL} & 425 & \num{2784} & \num{34318} & \num{804} \\
\isa{WITHIN} & 425 & \num{14622} & \num{199992} & \num{6165} \\
Masked \isa{NEAREST} & 425 & \num{12809} & \num{171711} & \num{5192} \\
\isa{REDUCE_OR} & 425 & \num{1746} & \num{25006} & \num{450} \\
\isa{REDUCE_AND} & 425 & \num{1746} & \num{25068} & \num{450} \\
\isa{ARGMAX_POP} & 425 & \num{12409} & \num{166711} & \num{5175} \\
\bottomrule
\end{tabular}
\end{table}

At this boundary \hades{} is faster than the ESP8266 by about $4.1$ (\isa{REDUCE_OR}, \isa{REDUCE_AND}) to $34$ (\isa{WITHIN}).

Against the $240$\,MHz ESP32 running the same kernels the factors are much smaller: $1.06$ for \isa{REDUCE_OR} and \isa{REDUCE_AND}, where the CPU does little per row, $1.9$ for \isa{EQUAL}, and up to $14.5$ for \isa{WITHIN}. The ESP32 is $2.4$ to $3.9$ times faster than the ESP8266 in time but comparable in cycles ($0.77$ to $1.26$ times), so its advantage is essentially its clock.

Against the $16$\,MHz eight-bit Arduino Nano, which runs all $255$ cases, the factors are $59$ to $471$. These ratios set the specified $40$\,MHz clock of \hades{} against microcontroller clocks of $16$, $80$ and $240$\,MHz.

On the ESP8266, scaling with the dataset size (runtime-$N$ engine) is approximately linear plus a fixed per-call overhead: the sixteen-row \isa{NEAREST} kernel takes $13.5$\,\textmu s and the $4096$-row kernel $3.23$\,ms, about $0.79$\,\textmu s per row. \hades{} has no compute-only figure beyond $N=16$. The tile-scan cycles for larger $N$ ($64$ tiles of $425$\,ns at $N=1024$) are derived and excluded from comparisons, because a real run must reload the data between tiles.

\subsection{Resident-query latency}\label{sec:br}

What does a query cost once the dataset and program are resident? Table~\ref{tab:br} gives the per-query cost with the dataset and program resident. On \hades{} the loop programs of Section~\ref{sec:templates} cost $24$ clocks for \isa{NEAREST} (four \isa{RECV}, a $17$-cycle \isa{SCAN}, two \isa{SEND}, one \isa{BRANCH}), $25$ for \isa{EQUAL}, $29$ for \isa{WITHIN}, $26$ for \isa{REDUCE_OR} and \isa{REDUCE_AND}, and $24$ for \isa{ARGMAX_POP}. The count is flat across datasets and seeds, equal at gate level for the cells covered by the gate-level runs, and, for masked similarity, the same for every mask density tested ($b=32,24,16,8,1$).

The scan is $17$ of the $24$ clocks for \isa{NEAREST}. The rest is operand and result transfer over the byte-wide port and the loop branch.

\begin{table}[t]
\centering
\caption{Resident-query (B-R) cost per query, $N=16$, uniform data. H: RTL cycles per query at $40$\,MHz. E, E32: measured steady-state time per query, runtime-$N$ engine ($B=1024$ for \isa{NEAREST}, \isa{EQUAL}, \isa{WITHIN}; $B=256$ for the masked and population workloads).}
\label{tab:br}
\footnotesize
\setlength{\tabcolsep}{3pt}
\begin{tabular}{@{}lrrrr@{}}
\toprule
Workload & H (cycles) & H (ns) & E (ns) & E32 (ns) \\
\midrule
\isa{NEAREST} & 24 & 600 & \num{16646} & \num{6005} \\
\isa{EQUAL} & 25 & 625 & \num{8354} & \num{2505} \\
\isa{WITHIN}, $T=0$ & 29 & 725 & \num{22745} & \num{7705} \\
\isa{WITHIN}, $T=14$ & 29 & 725 & \num{22997} & \num{7847} \\
Masked \isa{NEAREST}, $b=1$--$32$ & 24 & 600 & \num{16696} & \num{5935} \\
\isa{REDUCE_OR} & 26 & 650 & \num{6792} & \num{1414} \\
\isa{REDUCE_AND} & 26 & 650 & \num{6835} & \num{1410} \\
\isa{ARGMAX_POP} & 24 & 600 & \num{16086} & \num{5968} \\
\bottomrule
\end{tabular}
\end{table}

\begin{figure}[t]
\centering
\begin{tikzpicture}
\begin{axis}[
  xbar, y dir=reverse, width=0.72\columnwidth, height=7.6cm, bar width=3pt,
  xmin=0, xmax=29, ymin=-0.6, ymax=7.6,
  ytick={0,...,7}, yticklabels={{\texttt{NEAREST}}, {\texttt{EQUAL}}, {\texttt{WITHIN}, $T=0$}, {\texttt{WITHIN}, $T=14$}, {Masked \texttt{NEAREST}}, {\texttt{REDUCE\_OR}}, {\texttt{REDUCE\_AND}}, {\texttt{ARGMAX\_POP}}},
  tick label style={font=\scriptsize}, label style={font=\scriptsize},
  xlabel={Time per query (\textmu s)}, xmajorgrids, grid style={gray!25},
  tick align=outside, y tick style={draw=none},
  nodes near coords, nodes near coords align={horizontal},
  every node near coord/.append style={font=\tiny, /pgf/number format/.cd, fixed, precision=3},
  legend style={font=\scriptsize, at={(0.5,1.04)}, anchor=south, legend columns=2, draw=none, inner sep=1pt, column sep=3pt},
  legend image code/.code={\draw[#1] (0cm,-0.08cm) rectangle (0.25cm,0.08cm);},
]
\addplot[draw=black, fill=black!80] coordinates {(0.600,0) (0.625,1) (0.725,2) (0.725,3) (0.600,4) (0.650,5) (0.650,6) (0.600,7)};
\addplot[draw=black, fill=black!10] coordinates {(16.646,0) (8.354,1) (22.745,2) (22.997,3) (16.696,4) (6.792,5) (6.835,6) (16.086,7)};
\addplot[draw=black, fill=black!45] coordinates {(6.005,0) (2.505,1) (7.705,2) (7.847,3) (5.935,4) (1.414,5) (1.410,6) (5.968,7)};
\legend{{\hades{} (RTL, pre-silicon)}, {ESP8266 (measured)}, {ESP32 (measured)}}
\end{axis}
\end{tikzpicture}
\caption{Resident-query (B-R) time per query at $N=16$ for the workloads of Table~\ref{tab:br}. \hades{} (dark bars) is RTL cycles at the specified $40$\,MHz clock and is pre-silicon; the ESP8266 (light bars, $80$\,MHz) and the ESP32 (grey bars, $240$\,MHz) are measured. The Nano has no result at these batch sizes and is not shown.}
\label{fig:br}
\end{figure}
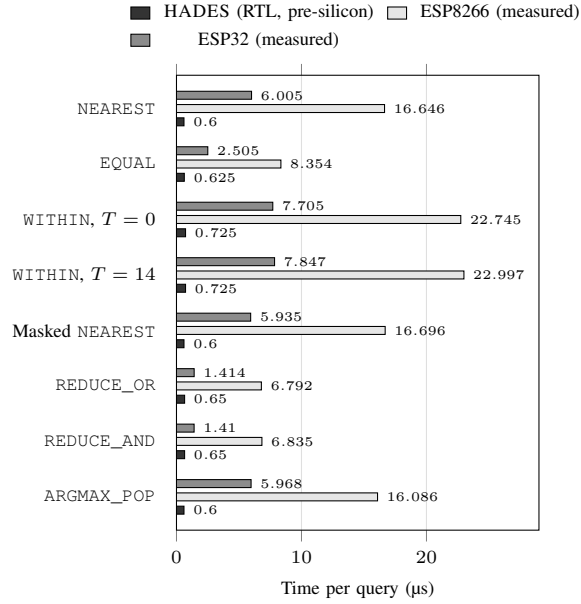

\hades{} is faster than the ESP8266 in every cell of Table~\ref{tab:br}, by about $10\times$ to $32\times$ (Figure~\ref{fig:br}), and faster than the ESP32 in every cell too, by about $2.2\times$ (\isa{REDUCE_OR}, \isa{REDUCE_AND}) to $11\times$ (\isa{WITHIN}). These factors are upper bounds, for the reason given below. \hades{} cycle counts show no data dependence in the cases tested: \isa{EQUAL} costs $25$ cycles at every hit density tested ($0$, $1/16$, $1/2$, $1$). A restart form of \isa{NEAREST}, which halts each query and restarts with \isa{EXECUTE}, costs $27$ clocks per query against $24$ for the loop form.

Table~\ref{tab:nano} gives the resident-query cost at $B=16$, the largest batch size on which the Nano runs, for all four platforms at that batch size. The Nano figures include the running hash, as the other boards do. The no-hash column removes it ($17$ to $45\%$ of the Nano's time in these rows). The Nano is an eight-bit part, so its factors ($225$ to $543$ in these rows) say little beyond the scale of that CPU. It also completes the mask-density and per-query-mask sets at $B=256$, which Table~\ref{tab:nano} omits.

\begin{table}[t]
\centering
\caption{Resident-query (B-R) cost per query at $N=16$, $B=16$, uniform data, in nanoseconds: the batch size at which the Nano runs. H: RTL cycles at $40$\,MHz. E, E32, N: measured, runtime-$N$ engine. N includes the running hash; N no-hash omits it (a Nano-only row).}
\label{tab:nano}
\footnotesize
\setlength{\tabcolsep}{3pt}
\begin{tabular}{@{}lrrrrr@{}}
\toprule
Workload & H & E & E32 & N & N no-hash \\
\midrule
\isa{NEAREST} & 600 & \num{16711} & \num{6021} & \num{245680} & \num{202945} \\
\isa{EQUAL} & 625 & \num{8440} & \num{2521} & \num{140754} & \num{76977} \\
\isa{WITHIN}, $T=0$ & 725 & \num{22870} & \num{7721} & \num{374047} & \num{247809} \\
\isa{WITHIN}, $T=14$ & 725 & \num{23129} & \num{7864} & \num{393687} & \num{267285} \\
\bottomrule
\end{tabular}
\end{table}

\paragraph{Serialisation and hashing in the baselines.} The microcontroller times of Tables~\ref{tab:br} and~\ref{tab:nano} include decoding the query, serialising the record and folding it into the running hash, none of which is part of the \hades{} boundary. On the Nano the hash alone is $17$ to $45\%$ of the per-query time (Table~\ref{tab:nano}); on the ESP boards no hash-free variant was measured, so the share is unknown and the factors of Table~\ref{tab:br} are upper bounds. For \isa{NEAREST} and \isa{EQUAL} a lower bound exists: the compute-only time of the same runtime-$N$ engine at $N=16$ (Section~\ref{sec:bc}) contains no decoding, serialisation or hashing, and Table~\ref{tab:hash} gives the corresponding factors. The B-C factors are lower bounds on the factor for a hash-free resident query, because such a query cannot take less time than the kernel alone, and the B-R factors are upper bounds, so for these two workloads the hash-free factor lies between them. No same-engine compute-only measurement exists for the other workloads, so only the upper bound applies to them.

\begin{table}[t]
\centering
\caption{Resident-query (B-R) and compute-only (B-C) times of the same runtime-$N$ engine at $N=16$, uniform data, in nanoseconds, and the factors relative to the \hades{} resident-query time ($600$\,ns for \isa{NEAREST}, $625$\,ns for \isa{EQUAL}; RTL cycles at $40$\,MHz). The compute-only time contains no decoding, serialisation or hashing, so its factor is a lower bound and the B-R factor an upper bound on the factor for a hash-free resident query. Factors are rounded to the nearest tenth. E, E32: measured.}
\label{tab:hash}
\footnotesize
\setlength{\tabcolsep}{3pt}
\begin{tabular}{@{}llrrrr@{}}
\toprule
Workload & Board & B-R (ns) & B-C (ns) & B-R factor & B-C factor \\
\midrule
\isa{NEAREST} & E & \num{16646} & \num{13501} & $27.7\times$ & $22.5\times$ \\
\isa{NEAREST} & E32 & \num{6005} & \num{5329} & $10.0\times$ & $8.9\times$ \\
\isa{EQUAL} & E & \num{8354} & \num{4413} & $13.4\times$ & $7.1\times$ \\
\isa{EQUAL} & E32 & \num{2505} & \num{1771} & $4.0\times$ & $2.8\times$ \\
\bottomrule
\end{tabular}
\end{table}

\paragraph{Dual-row and selective scans.} What do the scan variants buy in cycles? \isa{SCANP} gives the same result bytes as \isa{SCAN} and fewer cycles on the operations it accelerates (dual-row workloads). In the resident loop, \isa{EQUAL} drops from $25$ to $17$ clocks per query and \isa{REDUCE_OR} and \isa{REDUCE_AND} from $26$ to $18$. The saving of eight clocks equals the saving in the scan itself ($17$ to $9$ cycles). These \hades{}-internal comparisons of RTL cycle counts show what the added instruction buys in cycles, and Section~\ref{sec:impl} gives the cost in area. The baseline runs identical kernels for the \isa{SCAN} and \isa{SCANP} variants.

\isa{SCANSEL} is evaluated by the filter-and-refine workload, and Table~\ref{tab:s10} gives its cost against the selection size $k$. A sixteen-row \isa{NEAREST} refinement over the selected rows takes $1+k$ cycles, against $17$ for the full scan. The whole per-query program therefore costs $35+k$ clocks with a \isa{SCAN} first stage and $27+k$ with a \isa{SCANP} first stage, and both match the formula for every $k$ tested [RTL, GL]. The baseline runs the same two-stage kernel on the same rows and queries, and its time also grows with $k$.

\begin{table}[t]
\centering
\caption{Filter-and-refine (B-R, $N=16$, $B=64$) against selection size $k$. H: RTL clocks per query with a \isa{SCAN} or a \isa{SCANP} first stage. E: measured ns per query.}
\label{tab:s10}
\footnotesize
\setlength{\tabcolsep}{4pt}
\begin{tabular}{@{}rrrr@{}}
\toprule
$k$ & H \isa{SCAN} & H \isa{SCANP} & E (ns) \\
\midrule
0 & 35 & 27 & \num{7709} \\
1 & 36 & 28 & \num{8606} \\
2 & 37 & 29 & \num{9444} \\
4 & 39 & 31 & \num{11136} \\
8 & 43 & 35 & \num{14482} \\
16 & 51 & 43 & \num{21155} \\
\bottomrule
\end{tabular}
\end{table}

\paragraph{Structural data reduction.} A scan of the $16$ resident rows reads $64$ bytes locally and returns a short record. The ratio of local-read bytes to result bytes at $N=16$ is $32\times$ for \isa{NEAREST} ($64/2$), $21\times$ for \isa{EQUAL} ($64/3$), $10.7\times$ for \isa{WITHIN} ($64/6$), and $16\times$ for \isa{REDUCE_OR} and \isa{REDUCE_AND} ($64/4$). These ratios follow from the workloads and the record formats, and we did not measure them as speedups. They describe what leaves the memory interface of a near-memory engine and do not imply that the engine is faster than a host reading the same rows. We did not test whether a host pays a data-movement penalty at working sets that exceed its caches.

\subsection{End-to-end cost and data movement}
\label{sec:be}

Does the advantage survive host-side loading? End-to-end cost on \hades{} is dominated by loading. Three programs ($6$, $6$ and $4$ rows) load a $16$-row tile, each preceded by a reset and followed by \isa{EXECUTE}. This costs $478$ clocks per tile in the byte-serial loader protocol ($2$ clocks per loader byte, $1$ byte per clock on \isa{RECV}). Loading the resident loop program adds $50$ to $94$ clocks per tile, depending on the program length, which gives an initialisation cost of $528$ (\isa{NEAREST}), $532$ (\isa{EQUAL}), $548$ (\isa{WITHIN}) and $572$ (filter-and-refine) clocks per tile. Queries cost $24$ to $39$ clocks each in the cases shown here, plus two clocks to start. The tile-major order, which loads a tile once and runs all $B$ queries on it, is the primary \hades{} variant. The \hades{} cycle counts in this subsection are RTL cycles, are identical across five dataset seeds, and exclude host-side tile merging.

Table~\ref{tab:be} gives representative totals. For the microcontrollers, $T_{\mathrm{init}}$ is the time to deserialise the dataset from a byte buffer ($4.6$\,\textmu s at $N=16$ to $283$\,\textmu s at $N=1024$ on the ESP8266, $2.5$ to $158$\,\textmu s on the ESP32). Against the ESP8266, \hades{} has the lower total in $52$ of the $60$ cells of the end-to-end set ($4$ workloads, $5$ values of $N$, $B\in\{1,16,256\}$), by up to $30\times$. In the remaining eight cells (\isa{EQUAL} and filter-and-refine with $N\ge32$ at $B=1$) the ESP8266 finishes first, because the per-tile initialisation of \hades{} costs more than the ESP8266's per-dataset deserialisation and a single query does not amortise it. \isa{EQUAL} at $N=16$, $B=1$ is close to a tie ($14.0$ against $14.1$\,\textmu s).

Against the ESP32, which runs the same end-to-end set at $240$\,MHz, \hades{} has the lower total in $40$ of the $60$ cells, by up to $10\times$, and loses all $20$ cells with $B=1$, by up to $3.6\times$.

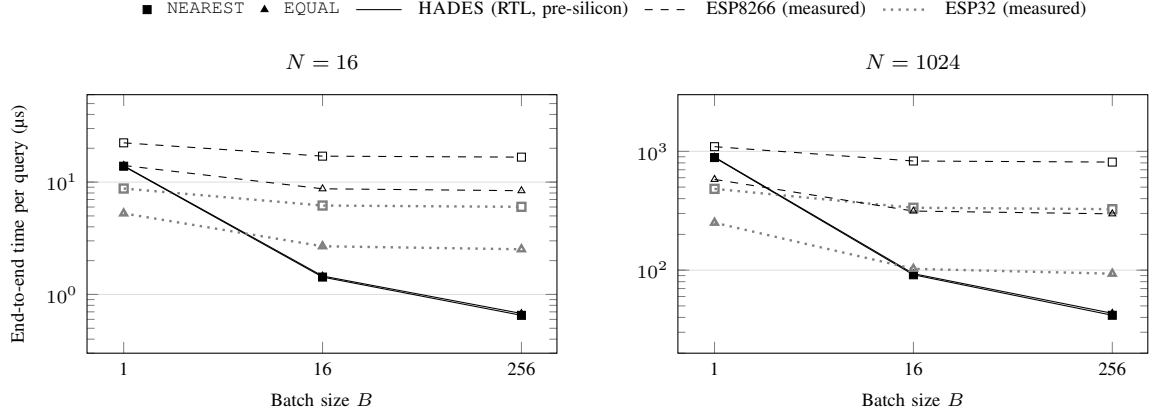
\begin{figure*}[t]
\centering
\begin{tikzpicture}
\begin{groupplot}[
  group style={group size=2 by 1, horizontal sep=1.6cm},
  width=0.43\textwidth, height=5cm, xmode=log, ymode=log,
  xtick={1,16,256}, xticklabels={1,16,256}, xmin=0.6, xmax=420,
  tick label style={font=\scriptsize}, label style={font=\scriptsize}, title style={font=\footnotesize},
  xlabel={Batch size $B$}, ymajorgrids, grid style={gray!25},
]
\nextgroupplot[title={$N=16$}, ymin=0.3, ymax=60, ylabel={End-to-end time per query (\textmu s)}, legend to name=fig:be:legend, legend columns=5, legend style={font=\scriptsize, draw=none, column sep=4pt}]
\addplot[black, solid, mark=square*, mark size=1.5pt, forget plot] coordinates {(1,13.85) (16,1.428) (256,0.6518)};
\addplot[black, dashed, mark=square, mark options={solid, fill=white}, mark size=1.5pt, forget plot] coordinates {(1,22.36) (16,17.02) (256,16.68)};
\addplot[black!50, dotted, line width=0.9pt, mark=square, mark options={solid, fill=white, draw=black!50}, mark size=1.5pt, forget plot] coordinates {(1,8.763) (16,6.185) (256,6.024)};
\addplot[black, solid, mark=triangle*, mark size=1.5pt, forget plot] coordinates {(1,13.97) (16,1.459) (256,0.6771)};
\addplot[black, dashed, mark=triangle, mark options={solid, fill=white}, mark size=1.5pt, forget plot] coordinates {(1,14.06) (16,8.717) (256,8.384)};
\addplot[black!50, dotted, line width=0.9pt, mark=triangle, mark options={solid, fill=white, draw=black!50}, mark size=1.5pt, forget plot] coordinates {(1,5.263) (16,2.685) (256,2.524)};
\addlegendimage{only marks, mark=square*, mark size=1.5pt}\addlegendentry{\texttt{NEAREST}}
\addlegendimage{only marks, mark=triangle*, mark size=1.5pt}\addlegendentry{\texttt{EQUAL}}
\addlegendimage{black, solid}\addlegendentry{\hades{} (RTL, pre-silicon)}
\addlegendimage{black, dashed}\addlegendentry{ESP8266 (measured)}
\addlegendimage{black!50, dotted, line width=0.9pt}\addlegendentry{ESP32 (measured)}
\nextgroupplot[title={$N=1024$}, ymin=20, ymax=3000]
\addplot[black, solid, mark=square*, mark size=1.5pt] coordinates {(1,886.4) (16,91.4) (256,41.71)};
\addplot[black, dashed, mark=square, mark options={solid, fill=white}, mark size=1.5pt] coordinates {(1,1095) (16,829.3) (256,812.4)};
\addplot[black!50, dotted, line width=0.9pt, mark=square, mark options={solid, fill=white, draw=black!50}, mark size=1.5pt] coordinates {(1,483.4) (16,335.1) (256,326)};
\addplot[black, solid, mark=triangle*, mark size=1.5pt] coordinates {(1,894.4) (16,93.4) (256,43.34)};
\addplot[black, dashed, mark=triangle, mark options={solid, fill=white}, mark size=1.5pt] coordinates {(1,580.5) (16,314.4) (256,297.8)};
\addplot[black!50, dotted, line width=0.9pt, mark=triangle, mark options={solid, fill=white, draw=black!50}, mark size=1.5pt] coordinates {(1,250.9) (16,102.6) (256,93.36)};
\end{groupplot}
\node[anchor=south] at ($(group c1r1.north east)!0.5!(group c2r1.north west)+(0,0.75cm)$) {\ref{fig:be:legend}};
\end{tikzpicture}
\caption{End-to-end (B-E) time per query against batch size $B$ for \isa{NEAREST} and \isa{EQUAL}, at $N=16$ (left) and $N=1024$ (right): uniform data, tile-major loading, ideal host, host-side merge excluded. Each point is the end-to-end total of the evaluated set divided by $B$, at the batch sizes of that set ($B\in\{1,16,256\}$); lines join the reported points only. Solid: \hades{}, RTL cycles at the specified $40$\,MHz clock (pre-silicon). Dashed: ESP8266, measured at $80$\,MHz. Dotted grey: ESP32, measured at $240$\,MHz. The initialisation cost of \hades{} is spread over the batch, whereas the per-query time of the microcontrollers stays near their resident-query times. At $B=1$ the ESP32 finishes first for both workloads, and the ESP8266 finishes first for \isa{EQUAL} at $N=1024$.}
\label{fig:be}
\end{figure*}

These counts hold for the evaluated conditions: \isa{NEAREST}, \isa{EQUAL} and \isa{WITHIN} at $T=14$, and filter-and-refine at $k=4$; uniform data (D1); tile-major loading; an ideal host; and host-side tile merge excluded. The comparison also counts asymmetrically: the baseline kernels are compiled for the measured platform, so no program-load time is counted for them, whereas the initialisation of \hades{} includes loading the loop program. Figure~\ref{fig:be} plots the end-to-end time per query against the batch size for $N=16$ and $N=1024$ for two of the four workloads; the others are in the set of Table~\ref{tab:be}.

\begin{table}[t]
\centering
\caption{End-to-end (B-E) totals for a batch of $B$ queries, in microseconds. H: RTL cycles $\times$ $25$\,ns, tile-major, ideal host, host merge excluded; the initialisation share is listed separately. E, E32: measured $T_{\mathrm{init}}+B\,T_q$.}
\label{tab:be}
\footnotesize
\setlength{\tabcolsep}{3pt}
\begin{tabular}{@{}lrrrr@{}}
\toprule
Workload, $N$, $B$ & H init & H total & E & E32 \\
\midrule
\isa{NEAREST}, 16, 1 & 13.2 & 13.9 & 22.4 & 8.76 \\
\isa{NEAREST}, 16, 256 & 13.2 & 167 & \num{4271} & \num{1542} \\
\isa{NEAREST}, 1024, 1 & 845 & 886 & \num{1095} & 483 \\
\isa{NEAREST}, 1024, 256 & 845 & \num{10678} & \num{207983} & \num{83446} \\
\isa{EQUAL}, 1024, 1 & 851 & 894 & 581 & 251 \\
\isa{WITHIN}, 1024, 256 & 877 & \num{12758} & \num{274537} & \num{107444} \\
Filter-and-refine, 1024, 1 & 915 & 981 & 646 & 319 \\
Filter-and-refine, 1024, 256 & 915 & \num{16893} & \num{93071} & \num{41230} \\
\bottomrule
\end{tabular}
\end{table}

A tile load costs about as many clocks as $22$ resident \isa{NEAREST} queries. Reloading every tile for every query (query-major) costs $9.7\times$ the tile-major total for \isa{NEAREST} at $B=16$ and $N=16$, $64$ and $256$ (for $N=16$, $8{,}880$ against $914$ clocks).

\paragraph{Host-transfer sensitivity.} The cycle counts above assume an ideal host. We simulate a slower host in RTL by inserting idle clocks after every \isa{RECV} byte of the resident loop. The cost follows $\text{base}+(\text{\isa{RECV} bytes}-1)\times\text{gap}$ clocks per query in all $18$ runs (Table~\ref{tab:gap}) [RTL]. A $15$-clock gap raises \isa{NEAREST} from $24$ to $69$ clocks.

\begin{table}[t]
\centering
\caption{Resident-loop clocks per query against idle clocks inserted after each \isa{RECV} byte (measured in RTL).}
\label{tab:gap}
\footnotesize
\setlength{\tabcolsep}{4pt}
\begin{tabular}{@{}lrrrrrr@{}}
\toprule
Gap (clocks) & 0 & 1 & 2 & 3 & 7 & 15 \\
\midrule
\isa{NEAREST} & 24 & 27 & 30 & 33 & 45 & 69 \\
\isa{EQUAL} & 25 & 28 & 31 & 34 & 46 & 70 \\
\isa{WITHIN} & 29 & 33 & 37 & 41 & 57 & 89 \\
\bottomrule
\end{tabular}
\end{table}

\paragraph{Link model.} The physical link is a model [MODEL]: the byte counts of a run (program, dataset, query and result bytes) are divided by an assumed half-duplex link rate. For $B=16$, the modelled link time divided by the chip time is $0.37$--$0.43$ at $40$\,MB/s, $1.5$--$1.7$ at $10$\,MB/s, $15$--$17$ at $1$\,MB/s and $150$--$170$ at $0.1$\,MB/s (\isa{NEAREST} to filter-and-refine). Over the evaluated range $N=16$ to $1024$ these ratios did not vary with $N$, since the chip time and the byte counts both grew with the number of tiles. In the model, the link time exceeds the chip time at $10$\,MB/s and below. This observation concerns the model, which assumes a half-duplex link with no protocol overhead.

\subsection{Gate-level confirmation}
\label{sec:gl}

Does the post-layout netlist behave like the RTL? The gate-level runs execute the instruction-cycle checks ($154$ items), the compute-only latency set ($255$ items, bytes and cycles), the resident-query set ($70$ items, every \texttt{READY} pulse cycle and byte) and the filter-and-refine set ($60$ items) on the post-layout netlist. All items are identical to RTL in bytes and in cycles [GL]. This agreement covers those item sets only, on a netlist hardened locally from the frozen sources whose cell count, cell area, macro area and worst slack equal those of the continuous-integration run. The project's continuous-integration gate-level job, which runs the \texttt{test/} cocotb modules on the netlist of that run, succeeds on the frozen sources; we did not rerun the full suite at gate level ourselves, and make no claim of full RTL-to-netlist equivalence.

\subsection{Energy}
\label{sec:energy}

\paragraph{Scope.} What do the measured baselines and the estimated \hades{} energy each tell us? Energy is measured at board level on three baseline microcontroller boards and estimated for \hades{}. These are different evidence classes with different scopes. We compute no energy ratio between them or between boards, and an energy advantage of \hades{} is not established.

\paragraph{Baseline measurement.} Each board is powered through an INA226 current monitor with a $0.1\,\Omega$ (R100) shunt in its supply path. The monitor converts shunt and bus voltage in $140$\,\textmu s each without averaging. A separate microcontroller reads it at more than $1$\,kS/s and integrates power over time. After the correctness gate, each board runs a fixed timeline from power-on with no synchronisation signal: a $20$\,s settling run of \isa{NEAREST}, then three repetitions of four $40$\,s phases. The phases are idle (a \texttt{delay} loop), a spin loop with the same batch structure and no kernel, \isa{NEAREST} and \isa{EQUAL}. The meter integrates the central $24$\,s of every phase. The kernels are those of the resident-query measurements at $N=16$, run back to back in batches of $1024$ queries ($16$ on the Nano, the largest batch size measured on that board).

The energy per query is the mean power of the kernel phase times the steady-state time per query of the same loop. That time comes from the separate timing runs of the same kernels (Tables~\ref{tab:br} and~\ref{tab:nano}) and is not counted during the energy run. We compute power from the raw shunt and bus voltages with the fitted $0.1\,\Omega$ shunt value and did not calibrate against a reference instrument, so absolute values carry the shunt tolerance and the monitor's gain and offset errors. For the kernel phases, the three repetitions of a phase differ (maximum minus minimum) by at most $0.04$\,mW on the ESP8266, $1.0$\,mW ($0.3\%$ of the mean) on the ESP32 and $0.8$\,mW ($0.8\%$) on the Nano.

\begin{table}[t]
\centering
\caption{Measured board power and energy per resident query, $N=16$ (MEAS-E). $P$: mean power in the kernel phase; spin: mean power in the spin phase; $t_q$: steady-state time per query of the same kernel from the timing runs of Tables~\ref{tab:br} and~\ref{tab:nano}; $E_q=P\,t_q$. Boundaries differ between boards (see text), as do clocks and batch sizes, so rows of different boards are not comparable.}
\label{tab:energy-meas}
\footnotesize
\setlength{\tabcolsep}{2.5pt}
\begin{tabular}{@{}llrrrr@{}}
\toprule
Board, $B$ & Workload & $P$ (mW) & Spin (mW) & $t_q$ (\textmu s) & $E_q$ (\textmu J) \\
\midrule
ESP8266, 1024 & \isa{NEAREST} & $74.9$ & $76.1$ & $16.65$ & $1.25$ \\
 & \isa{EQUAL} & $76.2$ & $76.1$ & $8.35$ & $0.637$ \\
Nano, 16 & \isa{NEAREST} & $97.6$ & $98.3$ & $245.7$ & $24.0$ \\
 & \isa{EQUAL} & $98.3$ & $98.3$ & $140.8$ & $13.8$ \\
ESP32, 1024 & \isa{NEAREST} & $296.7$ & $301.5$ & $6.00$ & $1.78$ \\
 & \isa{EQUAL} & $301.7$ & $301.5$ & $2.51$ & $0.756$ \\
\bottomrule
\end{tabular}
\end{table}

\paragraph{Marginal kernel energy.} On all three boards the kernel phase draws the same power as the spin loop to within $1.7\%$: \isa{NEAREST} is $0.7$ to $1.7\%$ below it and \isa{EQUAL} within $0.2\%$ of it. The marginal energy of a kernel over the spin loop, $(P-P_{\mathrm{spin}})\,t_q$, is therefore below $2\%$ of the energy per query, negative for \isa{NEAREST}, and unresolved. The \isa{NEAREST} shortfall exceeds the spread between repetitions, but the measurement does not show its cause, and we do not read it as a property of the kernel. The idle phase gives no floor for the kernel either: it draws $62.6$\,mW on the ESP8266 and $211.6$\,mW on the ESP32, whose \texttt{delay} sleeps, and $98.5$\,mW on the Nano, whose \texttt{delay} busy-waits.

Because kernel power is nearly constant, the energy per query on each board largely follows the time per query, and the ordering of the two workloads in energy follows their ordering in time (Table~\ref{tab:energy-meas}).

\paragraph{Measurement boundaries.} All values are whole-board supply measurements taken through the INA226 with USB unplugged, at different points. ESP8266: the $3.3$\,V module rail (about $3.5$\,V at the board), which bypasses the board's onboard regulator and USB-UART path. Wi-Fi is off, and a $1000$\,\textmu F capacitor on the measured $3.3$\,V supply prevented a startup brownout. ESP32: the board $5$\,V input ($4.4$\,V at the board under load), so the onboard regulator and USB-UART path lie inside the measured boundary. The radio is not started. Nano: the \texttt{5V} pin ($5.2$\,V), so the FT232 USB-serial interface is unpowered.

\paragraph{EST-M and EST-W.} Two estimates of \hades{} energy per resident query are available, and they are different evidence classes. EST-M covers only the \texttt{rf\_top} macro: the energy per cycle from circuit simulation of a slice of the extracted macro netlist (the shared logic and one and two of the $32$ bit columns, with ideal input sources and a lumped output load, $32$ measured cycles per access pattern), extrapolated linearly to $32$ bits ($17$\,pJ idle; $20$--$29$\,pJ for random two-read-one-write activity, the two values coming from two random access sequences; $34.7$\,pJ for an alternating worst-case pattern), multiplied by the RTL cycle count. EST-W covers the whole chip: the OpenSTA power of the post-layout standard cells, with switching activity captured from the gate-level simulation (run at a $20$\,ns clock, with the time axis of the activity file scaled to the $25$\,ns period before power analysis), plus the EST-M macro term. We accepted EST-W only after four checks fixed beforehand passed: all $8{,}669$ pin activities annotated; the analysis without activity reproducing the flow's power exactly; idle power below scan power with a consistent clock; and zero macro power in the standard-cell analysis.

\begin{table}[t]
\centering
\caption{Pre-silicon energy estimates per resident query, $N=16$ (nJ per query). EST-M: macro only. Standard cells: simulated activity at nominal TT, $1.8$\,V, $25^{\circ}$C. EST-W whole chip adds the EST-M range. The EST-M entry for \isa{WITHIN} is the generic $29$-cycle \isa{WITHIN} estimate; the standard-cell entry for \isa{WITHIN} corresponds specifically to the $T=14$ cell.}
\label{tab:energy}
\footnotesize
\setlength{\tabcolsep}{3pt}
\begin{tabular}{@{}lrrrr@{}}
\toprule
Workload & Cycles & EST-M & Std. cells & EST-W whole chip \\
\midrule
\isa{NEAREST} & 24 & 0.41--0.83 & 0.644 & 1.05--1.48 \\
\isa{EQUAL} & 25 & 0.43--0.87 & 0.668 & 1.09--1.54 \\
\isa{WITHIN} & 29 & 0.49--1.01 & 0.797 & 1.29--1.80 \\
\bottomrule
\end{tabular}
\end{table}

Table~\ref{tab:energy} gives the estimates. The idle standard-cell power is $0.781$\,mW and the scan loops draw $1.07$--$1.10$\,mW, so the clock tree and sequential cells dominate and the energy per query is mostly background power multiplied by the query time. The increment over idle is $0.175$, $0.180$ and $0.230$\,nJ per query for \isa{NEAREST}, \isa{EQUAL} and \isa{WITHIN}.

The activity comes from a unit-delay gate-level simulation, which includes glitches that real delays would change, and from one dataset and $256$ queries per workload. The estimates exclude pad and input/output energy, the loader and the chip-to-host link. The macro term has no library power model, and it carries a pattern-dependent uncertainty: the two random access sequences differ by about $40\%$, and the slice extrapolation is not checked against a full-macro simulation. The estimates hold at one operating corner.

\subsection{Sensitivity}
\label{sec:sens}

\paragraph{HADES clock.} The $40$\,MHz clock is the specification that timing analysis closes. Converting the same cycle counts at $20$ and $10$\,MHz, the \isa{NEAREST} resident query takes $1{,}200$ and $2{,}400$\,ns against $600$\,ns at $40$\,MHz, and the sixteen-row scan $850$ and $1{,}700$\,ns against $425$\,ns. This conversion of cycles to time adds no measurement.

\paragraph{Baseline clock and optimisation.} On the ESP8266, $160$\,MHz halves the time exactly (\isa{NEAREST} $13{,}501$\,ns at $N=16$ becomes $6{,}750$\,ns, and $807{,}381$\,ns at $N=1024$ becomes $403{,}670$\,ns), as observed in all $48$ compute-only groups measured at both clocks. \texttt{-O3} stays within about $2\%$ of \texttt{-O2}, because the library popcount routine dominates. \hades{} remains faster in the cells measured at $160$\,MHz (compute-only \isa{NEAREST} and \isa{EQUAL}, and resident-query \isa{NEAREST}, \isa{EQUAL} and \isa{WITHIN}), with roughly half the $80$\,MHz ratio factors. The end-to-end set was not repeated at $160$\,MHz, so the end-to-end comparison is untested at that clock. These rows do not test hand-optimised baselines: a SIMD or table-driven popcount could change the baseline times, and we did not run that experiment.

\section{Discussion}
\label{sec:discussion}

\subsection{What the evidence supports}

\textbf{A small substrate suffices at this scale.} For the research question, the evidence shows that twelve opcodes, three working registers, a $32$-slot program and one shared register-file macro expressed all ten workloads of Section~\ref{sec:workloads}, so control and storage did not limit them. Program storage added no macro of its own, because the program lives in the macro that holds the data. The part of the design that grows with workload variety is the sequencing logic and the instruction slots, and the datapath stays fixed: \isa{SCANP} and \isa{SCANSEL} leave the popcount datapath unchanged, and their area comes from a second-row consumer, a selection sequencer and their control.

The work does not establish a minimum. We ran no ablation on the opcode set, the number of program slots, the register widths or the row capacity, so the result is one sufficient point in the design space. The slot counts bound the question from one side only: $16$ slots would not hold the longest template, and no workload in the set needs a depth between $19$ and $32$.

\textbf{The interface shapes the program.} In the resident loops $7$ to $12$ of the $24$ to $29$ clocks per query are \isa{RECV}, \isa{SEND} and the loop branch, so the narrow interface shows up in the program size and in the per-query time as well as in the loading time.

Physically, the standard cells ($26{,}968$\,\textmu m$^2$) are about $1.7$ times the area of the $15{,}744$\,\textmu m$^2$ macro, although the datapath is a masked XOR and a population count. Multi-input combinational logic is about $56\%$ of the standard-cell area, sequential cells about $21\%$, and timing-repair buffers about $15\%$. The area is not broken down by function, so how much of that logic is control, loader, result registers or datapath is not resolved, and the cost of programmability as a fraction of the design is not quantified.

\paragraph{Cost and benefit of the scan variants.} Against the base ISA (Table~\ref{tab:cost}), \isa{SCANP} and \isa{SCANSEL} together cost $17.4\%$ more standard-cell area and $25\%$ more routed wirelength, with $1.29$\,ns less setup slack in a single comparison (Section~\ref{sec:impl}), and they buy fewer cycles on the operations they accelerate (Section~\ref{sec:eval}). The value of that saving depends on the boundary. Subtracting the eight clocks per query from the end-to-end totals (a derivation from the reported totals), the saving is $8$ of \isa{EQUAL}'s $559$ clocks ($1.4\%$) at $N=16$ and $B=1$ and $2{,}048$ of its $6{,}934$ clocks ($30\%$) at $B=256$. The variants pay off when queries repeat over resident data, the regime they were designed for, and matter little for a single query that must also load its dataset.

\paragraph{Capacity and loading.} The $16$-row capacity and the loading path set what a resident dataset means for this design. A program writes a row with four \isa{RECV} and one \isa{STA} (five slots, ten loader bytes for four data bytes). For the four end-to-end workloads, loading is about $93$ to $95\%$ of the clocks of a run with $B=1$ and $5$ to $8\%$ with $B=256$, and these shares do not depend on $N$ because both terms grow with the number of tiles. The host must tile and reload a dataset of more than $16$ rows, and the query-major order that reloads for every query costs $9.7\times$ the tile-major order. A dedicated data-load path would change this trade-off, but it is a different design that we did not evaluate.

\subsection{Boundary and baseline dependence}

The ordering between \hades{} and the microcontrollers depends on the baseline, the boundary and the batch size (Section~\ref{sec:eval}). At the compute-only and resident-query boundaries \hades{} is faster than both the ESP8266 and the ESP32, with factors that fall steeply against the $240$\,MHz ESP32 because the ESP32 advantage over the ESP8266 is mostly its clock. The resident-query factors are upper bounds because the baseline times include record hashing (Table~\ref{tab:hash}). The Nano, which runs only a subset, has longer times than either.

\hades{} amortises its initialisation over a batch, and every end-to-end cell it loses has $B=1$ (Section~\ref{sec:be}). The end-to-end cells for \isa{EQUAL} and filter-and-refine at $B=1$ already favour the microcontroller, because per-tile loading exceeds its per-dataset deserialisation, and the modelled link time exceeds the chip time at $10$\,MB/s and below. Outside the scan itself, the interface (about $29$ to $41\%$ of each resident query) and the load path (most of a single end-to-end query) are the dominant costs at this scale.

These results hold for plain C on these microcontrollers. They do not establish an advantage over tuned microcontroller code, over the ESP8266 at $160$\,MHz for end-to-end runs (which we did not repeat there), over a higher-performance host, or at the system level. They also do not show that an ESP8266 driving \hades{} would see these times.

\subsection{Implications for processing-near-memory design}

Three points follow from the experiments, each limited to this design and these workloads.

A program-driven engine at this scale can be assembled around an unmodified register-file macro. What the programmable substrate adds around the datapath is sequencing, selection and byte-level I/O, and the measured cost of the two scan variants came from such additions.

The value of data reduction at the interface depends on where the interface cost falls. For queries over resident data the byte-wide port costs about $29$ to $41\%$ of each query, and loading costs most of a single query, so the engine suits repeated queries over a small resident set. This statement concerns a byte-wide, $40$\,MHz interface.

Evaluating at several boundaries and against two baselines changed what can be said: the boundary, the baseline clock and the batch size each alter the ordering, and a compute-only figure alone would have hidden both the interface cost and the load cost.

\paragraph{Energy.} The estimates give $1.05$ to $1.80$\,nJ per resident query for the whole chip (\isa{NEAREST}, \isa{EQUAL} and \isa{WITHIN}), of which $0.41$ to $1.01$\,nJ is the macro alone. Most of the standard-cell energy per query is idle power multiplied by the query time: the idle standard-cell power is $0.781$\,mW against $1.07$ to $1.10$\,mW during scans, so the increment over idle is only about $27$ to $29\%$ of the standard-cell energy. If that held in silicon, shortening a query in cycles would matter more for energy than reducing switching activity, but this is an inference from simulation.

On the boards a kernel draws within about $2\%$ of the power of a spin loop (Table~\ref{tab:energy-meas}), so the energy per query largely follows the execution time. This concerns the baselines and says nothing about \hades{}, though it is consistent with background power and query time dominating energy per query on both sides. The scopes differ (a whole board against a chip without pads and link, at different supply voltages and technologies), so we do not compare the estimates with the measurements, and the paper establishes nothing about energy relative to any other platform.

\section{Limitations and Threats to Validity}
\label{sec:limitations}

\paragraph{Evidence boundary.} Every \hades{} cycle count comes from RTL or gate-level simulation, and the $25$\,ns cycle is a specification that timing analysis closes. We did not measure the Tiny Tapeout board interface or the behaviour of a real host and link. The host in the experiments is ideal (Section~\ref{sec:eval}), and the link time is a model with an assumed half-duplex rate and no protocol overhead. The energy of \hades{} is an estimate, and neither estimate predicts silicon. No silicon exists.

\paragraph{Baselines.} The baseline kernels are plain C compiled once, with the compiler's own popcount, on the ESP8266, the ESP32 and the Nano. We did not run hand-optimised, SIMD, table-driven or code-placement variants, which could narrow or widen the gaps. The baseline results use two kernel engines whose results for the same vectors can differ, so comparisons are valid only within one engine. The results are for one board of each type, and the resident-query and end-to-end per-query times include record serialisation and hashing, whose share is unmeasured on the ESP boards ($17$ to $45\%$ on the Nano; Table~\ref{tab:hash} bounds it for two workloads). The Nano runs only part of the evaluation because of its RAM and flash, and the ESP32 timed regions run with interrupts masked and execute from flash through the instruction cache. We include no higher-performance host baseline, and the experiments do not demonstrate a host-memory data-movement penalty that near-memory execution avoids.

\paragraph{Energy measurement.} Power and time per query come from separate runs, and the energy loops were not instrumented, so we would not see a change in per-query speed between the two runs. Absolute values depend on the shunt tolerance and the monitor's gain and offset, and we did not calibrate against a reference instrument. Each value is the mean of three $24$\,s windows. The board boundaries differ (Section~\ref{sec:energy}), and kernel power equals spin power within $2\%$, so the marginal energy of a kernel is not resolved.

\paragraph{Scope of the workloads.} Rows are $32$ bits, the resident capacity is $16$ rows, and the workloads are the eight scan operations and the two scan variants. We evaluate larger datasets through a host-side tile-major extension, with an evaluation-side record format for $N>16$ that the workload definitions for $N=16$ do not specify. The host-side merge is excluded from chip time, and the query-major order runs only as a worst case. The evaluation fixed several parameters without sweeping them: the $B=256$ batch for the mask-density and per-query-mask sets, the \isa{WITHIN} threshold ($T=14$) and filter-and-refine selection size ($k=4$) in the end-to-end set, which uses uniform data only, and the purpose-built dataset behind the filter-and-refine selection.

\paragraph{Verification and implementation.} Gate-level agreement covers only the item sets of Section~\ref{sec:gl}, on a netlist hardened locally from the frozen sources with the same flow; its cell count, cell area, macro area and worst slack equal those of the project's continuous-integration run, but the two netlists were not compared directly. Simulation and hardening used the SystemVerilog generated from the Veryl sources, and we did not independently recheck the correspondence between the two. The physical result is a single hardening run per design, so run-to-run variation is not characterised and the costs of \isa{SCANP} and \isa{SCANSEL} cannot be separated.

\paragraph{Architectural constraints.} The architecture has no hardware flow control between \isa{SEND} and \isa{RECV}, no \isa{RECV} timeout, no support for reloading the program while it executes, and undefined behaviour for the reserved opcodes. These are limits of the design as described, and none of the experiments stresses them beyond the documented host protocol.

\section{Conclusion}
\label{sec:conclusion}

\hades{} is a small program-driven processing-near-memory engine for binary similarity search. Its design follows from the workload: a scan reads every resident row and returns a short record, so a masked XOR and a population count suffice as a datapath, and the remaining design question is how little control and storage make that datapath programmable. Twelve $16$-bit opcodes, three working registers and a $32$-slot program, built around one shared register-file macro, express ten workloads in $8$ to $19$ slots with byte-exact results. \isa{SCANP} uses the second read port to reduce the cycle count of equality and reduction scans, and \isa{SCANSEL} keeps filter-then-refine inside the engine, in the \hades{} cycle model. The design is hardened on a $2\times2$ Tiny Tapeout tile with timing closed at $40$\,MHz and the standard cells occupying $26{,}968$\,\textmu m$^2$ beside the macro. This is a post-layout result.

The result depends on the measurement boundary. At the resident-query boundary \hades{} is faster than both microcontrollers, by $10\times$ to $32\times$ against the ESP8266 and $2.2\times$ to $11\times$ against the ESP32; these factors are upper bounds, because the baseline times include record hashing, and the compute-only time of the same engine gives lower bounds of $22.5\times$ and $7.1\times$ (ESP8266) and $8.9\times$ and $2.8\times$ (ESP32) for \isa{NEAREST} and \isa{EQUAL}. End to end, the outcome turns on batch size. In the evaluated set \hades{} has the lower total at both $B=16$ and $B=256$ against both boards, and it loses single-query runs: all of them to the ESP32 and eight to the ESP8266. Loading the dataset and program is about $93$ to $95\%$ of the clocks of a single-query run. Against the ESP32 a single query never repays that cost, against the ESP8266 it repays it in $12$ of the $20$ single-query cells, and a larger batch repays it against both. These results hold for plain C baselines under the evaluated conditions and do not extend to tuned code or to a higher-performance host.

The energy results are narrower. On all three microcontrollers kernel power stays within about $2\%$ of a spin loop, so query energy largely follows query execution time at the measured boundaries. The energy of \hades{} is estimated.

Architecturally, the lesson is that the programmable control and instruction storage fit around a shared memory macro, and that the byte-wide interface and the loading path then account for a large part of the cost of a query. The constraint that shaped the design is the interface, not the datapath. What remains open is whether programmability stays this cheap in silicon, how the engine compares with tuned baselines, and how a data path that does not pass through the program stream would change the loading cost.

\bibliographystyle{IEEEtran}
\bibliography{references}

\end{document}